\documentclass[12pt,letter,english,intoc,bibliography=totoc,index=totoc,BCOR10mm,captions=tableheading,nottitlepage,fleqn]{article}

\usepackage{setspace, geometry}
\usepackage{etex}
\usepackage{babel,newcent,textcomp}

\usepackage{scrextend}
\usepackage{booktabs, threeparttable}

\usepackage[T1]{fontenc}
\usepackage[latin9]{inputenc}
\usepackage[autostyle]{csquotes} 
\usepackage[style=bwl-FU, citestyle=authoryear, backend=biber, useprefix, maxbibnames=99]{biblatex}
\usepackage{lmodern}

\usepackage{sgame}
\usepackage{fancyhdr}
\usepackage{amsmath,amssymb,bbm}
\usepackage[amsthm]{ntheorem}
\usepackage{dsfont}
\usepackage{tablefootnote}
\usepackage[labelfont=bf]{caption}
\usepackage{subcaption}
\usepackage{enumitem}
\usepackage{lscape}

\usepackage{tikz}

\usetikzlibrary{shapes.geometric, arrows, positioning}
\tikzset{
    >=stealth',
    box/.style={
           rectangle,
           rounded corners,
           draw=black, very thick,
           text width=6.5em,
           minimum height=1.5em,
           text centered},
    boxsmall/.style={
           rectangle,
           rounded corners,
           draw=black, very thick,
           text width=6em,
           minimum height=1.5em,
           text centered},
     simple/.style={
           text centered},
    arrow/.style={
           ->,
           thick,
           shorten <=2pt,
           shorten >=2pt,}
}
\usepackage{graphicx}
\graphicspath{{""}}

\usepackage[multiple]{footmisc}
\usepackage[unicode=true,
 bookmarks=true,bookmarksnumbered=true,bookmarksopen=true,bookmarksopenlevel=1,
 breaklinks=false,pdfborder={0 0 0},backref=false,colorlinks=true,linkcolor=black,citecolor=black,anchorcolor=black,urlcolor=blue]
 {hyperref}
\hypersetup{pdftitle={Your title},
 pdfauthor={Your name},
 pdfpagelayout=OneColumn, pdfnewwindow=true, pdfstartview=XYZ, plainpages=false}

\usepackage{cleveref}
\usepackage[figure]{hypcap}
\usepackage{calc}

\let\mySection\section\renewcommand{\section}{\suppressfloats[t]\mySection}

\theoremstyle{break}

\newtheorem{hypothesis}{Hypothesis}

\newtheorem*{fact-non}[theorem]{Fact}

\makeatother

\begin{document}

\lhead[]{}

\rhead[]{}

\lfoot[\thepage]{}

\cfoot{}

\rfoot[]{\thepage}

\newpage

\title{Good News Is Not a Sufficient Condition for \\ Motivated Reasoning\thanks{\scriptsize{This paper was previously circulated with the title: ``Do People Engage in Motivated Reasoning to Think the World Is a Good Place for Others?'' I would like to thank Alberto Alesina, Roland Benabou, Christine Exley, Amanda Friedenberg, David Laibson, Matthew Rabin, Mattie Toma, Leeat Yariv, and seminar participants at Harvard and the ESA Global Meetings for helpful comments. I am grateful for funding support from Harvard Business School and the Eric M. Mindich Research Fund for the Foundations of Human Behavior. Experiments were conducted when I was at Harvard, and were deemed exempt by the IRB at Harvard (IRB17-1725).}}}
\author{Michael Thaler\thanks{\scriptsize{University College London. Email: \href{michael.thaler@ucl.ac.uk}{michael.thaler@ucl.ac.uk}.}}}
\date{January 2024}

\maketitle

\abstract{People often receive good news that makes them feel better about the world around them, or bad news that makes them feel worse about it. This paper studies how the valence of news affects belief updating, absent functional and ego-relevant factors. Using experiments with over 1,500 participants and 5,600 observations, I test whether people engage in motivated reasoning to overly trust good news versus bad news on valence-relevant issues like cancer survival rates, others' happiness, and infant mortality. The estimate for motivated reasoning towards good news is a precisely-estimated null. Modest effects, of one-third the size of motivated reasoning in politics and performance, can be ruled out. Complementary survey evidence shows that most people expect good news to increase happiness, but to not systematically lead to motivated reasoning. These results suggest that belief-based utility is not sufficient in leading people to distort belief updating in order to favor those beliefs. 

\vspace{7mm}
\noindent \textbf{JEL classification:} C91; D83; D91
\vspace{0mm}

\noindent \textbf{Keywords:} motivated reasoning; belief-based utility; experimental economics

\setcounter{page}{0}
\thispagestyle{empty}

}




\newpage

\section{Introduction}

When people receive information with positive or negative valence, it often evokes an emotional response. Valence is a common attribute of news. For instance, people may receive ``good news'' that reports that there have been improvements in reducing childhood cancer mortality rates around the world, or ``bad news'' that reports the decreasing level of Americans' life satisfaction. The former leads them to feel better about the world and the second leads them to feel worse about it. Such emotions may distort information processing by leading people to overly trust good news more, and bad news less, because they seek to maintain more optimistic beliefs. 

\textit{Motivated reasoning} provides one account of how people use emotions in inference. It posits that people systematically deviate from Bayesian updating when they receive new information, and form posterior beliefs that align with beliefs they find more attractive to hold. There is extensive evidence that motivated reasoning leads people to become overly optimistic about themselves and their identity in a variety of domains.\footnote{For instance, motivated reasoning has been shown in the domains of ability (\cite{MNNR22}; \cite{ER11}), altruism (\cite{E16}; \cite{EK-WP}; \cite{DPBS15}), attractiveness (\cite{ER11}), and politics (\cite{TL06}; \cite{T-WP}).} The literature suggests two competing explanations for the prevalence of motivated reasoning in such settings. One explanation is that motivated reasoning is driven by valence: people trust news more if it sends messages that they feel happier believing are true. This is a common explanation in the literature, but most examples focus on motives that are about oneself, rather than motives driven by positive and negative emotions about others. For instance, \citeauthor{W80}'s (1980) seminal paper defines ``optimism'' as being about one's future life outcomes relative to others; \textcite{BT16} provide examples of affective optimism about one's intelligence, attractiveness, and future well-being; and \textcite{BP05} provide examples of optimism about one's future income. 

A second explanation is that motivated reasoning is driven by the functional value of beliefs in affecting behavior, self-image, and persuasion: people form beliefs that help them be perceived as good and persuade others that their arguments are correct (e.g. \cite{BT02}; \cite{BT16}; \cite{STvdW-WP}). These two explanations often overlap, since optimism about oneself and one's identity may play a functional role in interactions with others (\cite{vHT11}; \cite{T11}; and \cite{SvdW19}). 

In this paper, I disentangle the valence and functional mechanisms by studying motivated reasoning in environments where news varies in its valence independently from its functional value, focusing on how people reason about \textit{others}. First, I consider questions for which valence varies but function plays a minimal role. For example, while a reduction in childhood cancer mortality rates would be ``good news'' for valence reasons, it has limited functional value. Second, I consider questions that vary in \textit{both} valence and function by varying their political ideology. For example, while high murder and manslaughter rates during the Obama administration is ``bad news'' for valence reasons, it is functionally useful for Republicans whose political identity is consistent with Obama's policies being ineffective at reducing crime. I test for motivated reasoning experimentally with over 5,600 observations from 1,500 participants in two waves, studying five topics for which valence is hypothesized to vary absent politics, and six topics for which valence and politics both vary. (Topics are listed in Appendix \Cref{topics-positivity}.)

To measure motivated reasoning, I adapt a design developed in a companion paper (\cite{T-WP}). he experimental design has two main steps: First, participants report the median of their belief distribution about factual questions (such as cancer survival rates). Second, participants assess the truthfulness of information sources that either tell them the median of their belief distribution is too high or too low. Because the median is elicited, Bayesians would infer nothing about the truthfulness of the source from the message they see. However, motivated reasoning often leads people to trust the news more if it aligns with the beliefs they are more motivated to hold. If motivated reasoning affects inference because of valence, then participants will rate good news as more likely to be true than bad news. In \textcite{T-WP}, I focus on identifying motivated reasoning in political and performance settings, but do not study valence. While there is some overlap in topics between the two papers, the hypotheses and experiments in this paper are novel.\footnote{The only data overlap is that this paper's political/performance questions are used in the replication exercise in Online Appendix D of \textcite{T-WP}.} 

The main result is that good news has a \textit{precisely-estimated null effect} on motivated reasoning. Overall, participants rate good news to be 0.2 pp (s.e.\ 0.7 pp) less likely to be true than bad news. In addition, the distribution of beliefs are indistinguishable, and the effect of valence is the same when function does not play a role and when it plays an independent role. To demonstrate the precision of these results, I also compare motivated reasoning about valence to other domains. Consistent with previous evidence, motivated reasoning about politics and performance is significant: participants rate news to be 5.3 pp (s.e.\ 0.8 pp) more likely to be true if it aligns with their political party's stances, and participants rate news to be 8.5 pp (s.e.\ 2.7 pp) more likely to be true if it tells them they performed better than expected. The effects of politics and performance are statistically-significantly larger than the effect of valence. These results indicate that it is not the experimental design that is leading to the null effect of valence, but rather that motivated reasoning is not driven by good news in the same way that it is in other settings. 

The aggregate null effect may be explained by good news not evoking motivated reasoning, or by there being equal shares of ``optimistic'' and ``pessimistic'' types in the sample. I show that the data align more with the first explanation. First, I use the initial guesses of participants to classify participants as optimistic or pessimistic on non-political questions.\footnote{A slight majority of initial guesses are pessimistic, indicating that there is no general optimism bias. This result differs markedly from the classic results from \textcite{W80} that show excessive optimism about one's own future prospects.} Optimists and pessimists assess good news and bad news the same way. This result contrasts with that of politics and performance, where priors are significantly correlated with news assessments. Second, demographics are also not predictive of valence-driven motivated reasoning.\footnote{The gender results are different than papers such as \textcite{T21}; \textcite{CCK-WP}; and \textcite{E11}, which find significant gender differences in motivated reasoning about performance or knowledge.} 

Robustness checks confirm these results. I also ask participants to update their median beliefs given the news, and valence does not significantly affect this belief updating. They are equally likely to revise their beliefs when receiving good news or bad news. Restricting to non-political (or political) questions yields the same null effects. Other specifications and sample restrictions do not meaningfully affect the results either.

I next consider two explanations for the main findings: Either participants do not see the valence questions as evoking positive and negative emotions, or these emotions are not a driver of motivated reasoning. As emotions are hard to directly measure, I use two surveys to ask people about their beliefs about the effects of valence on happiness and belief updating. These surveys provide evidence in favor of the second explanation. In Survey 1, respondents are asked to predict the role of motivated reasoning on issues that evoke good/bad news, politics, and performance. Similar shares of respondents expect there to be motivated reasoning towards good news and towards bad news, while significant majorities of respondents expect others to engage in pro-party and pro-performance motivated reasoning. These responses are well-aligned with the experimental results. Meanwhile, in Survey 2, a sizeable majority of respondents expect \textit{all three} categories of ``good news'' (valence, politics, and performance) to make people happier. That is, respondents expect valence to systematically affect happiness, but not affect motivated reasoning. These results indicate that respondents expect belief-based happiness to be insufficient for inducing motivated reasoning. 

Taking the experimental and survey results as a whole, these results are consistent with the notion that, absent functional factors, motivated reasoning is not well-explained by belief-based happiness. People may attain greater happiness by receiving good news and yet not systematically distort their inference process to favor these beliefs; conversely, the beliefs that people find more attractive to hold are not necessarily the ones that make them happier. Rather, other factors are necessary to drive motivated-reasoning biases. 

This paper contributes to the growing literature that studies which types of motives are sufficient for motivated reasoning by showing where motivated reasoning has its limits. Experimentally, this literature has shown that people form beliefs to help them win arguments and persuade others (\cite{SvdW19}; \cite{STvdW-WP}; \cite{SKPvH20}), defend their political identity (\cite{TL06}; \cite{T-WP}), and overstate their own intelligence (\cite{MNNR22}; \cite{ER11}), altruism (\cite{DWK07}; \cite{E16}), and future prospects (\cite{W80}; \cite{OSD13}; \cite{S11}). 



From an applied perspective, these results may help explain why motivated reasoning has been documented in some domains (\cite{K90}; \cite{TL06}; \cite{MNNR22}; \cite{ER11}) but not in others. A few previous papers have studied specific settings where motivated reasoning has not been detected: \textcite{SBLS-WP} find prior-driven asymmetric updating that leads to different updating processes about climate change, and \textcite{B20} find that monetary stakes may not be sufficient for motivated reasoning. In \textcite{T21}, I also show that gender affects the domains of motivated reasoning. There is a broader debate about the prevalence of motivated reasoning in general (\cite{B19}), as certain experimental designs find limited evidence of the bias (\cite{C18}; \cite{PR19}). This paper contributes to this literature by confirming the existence of motivated reasoning but unpacking one reason why there are different effects across domains. Given the focus of the topics studied, this paper also relates to the large literature on emotions and happiness in economic decision-making (e.g. \cite{L00}; \cite{FS02}; \cite{LLVK15}), and cautions against assuming a link between emotional cues like happiness and information processing. 

The rest of the paper proceeds as follows: \Cref{theory-design} discusses the experimental design and details. \Cref{results-main} presents the main results. \Cref{discussion} discusses interpretations of the results and presents the survey evidence. \Cref{conclusion} concludes and proposes directions for future work. The appendices provide additional results and show the exact questions and pages that participants see.

\section{Experiment Design}
\label{theory-design}

\subsection{Overview}

In order to identify motivated reasoning, I use an experiment that is designed to give participants information that would not affect a Bayesian's assessments but would affect a motivated reasoner's. I extend this design (discussed in complementary work: 
\cite{T-WP}) to questions in which valence varies. To fix ideas, consider the following question, taken verbatim from the experiment: 

\vspace{5mm}
\begin{addmargin}[1cm]{1cm}
\textit{Acute Myeloid Leukemia (AML) is a devastating illness in which cancerous cells emerge in the bone marrow, invade the blood stream, and may spread to the rest of the body. Tragically, hundreds to thousands of children under the age of 15 are diagnosed with AML each year; it is one of the most common cancers among children.}

\textit{Of children under the age of 15 who are diagnosed with AML, what percent survive for at least 5 years?}
\end{addmargin}
\vspace{5mm}

\noindent This is a question for which higher-valued states are coded as ``good'' but are coded as not affecting functional or ego-relevant motives. 

The main test of motivated reasoning involves three steps:

\begin{enumerate}
\item \textbf{Questions:} Participants are asked to guess the answers to questions like the one above. Importantly, they are asked and incentivized to guess their \textit{median belief} (i.e. such that they find it equally likely for the answer to be above or below their guess). The median is incentivized using a linear scoring rule.\footnote{Participants are also asked to give a confidence interval which is incentivized using piecewise-linear scoring rules, but I do not use this data in this paper.} 

\item \textbf{News text:} Participants receive a binary message from one of two randomly-chosen news sources: True News and Fake News. The message from True News is always correct, and the message from Fake News is always incorrect. This is the main (within-person) treatment variation.

\hspace{5mm} The message says either ``The answer is \textbf{greater than} your previous guess of [previous guess].'' or ``The answer is \textbf{less than} your previous guess of [previous guess].'' Note that the exact messages are \textit{different} for each participant since participants have different median guesses.

\hspace{5mm} For the question above, ``greater than'' is coded as good news and ``less than'' as bad news (\Cref{topics-positivity}). 

\item \textbf{News assessment:} After receiving the message, participants assess the probability that the message came from True News using a scale from 0/10 to 10/10, and are incentivized to state their true belief using a quadratic scoring rule. This news veracity assessment is the main outcome measure. The effect of variation in news direction on veracity assessments is the primary focus for much of this paper. Participants are also asked to give an updated median guess after seeing the message (again incentivized with a linear scoring rule). For details, see the Online Appendix.
\end{enumerate}

Because the participant has previously stated their median belief, $\mu$, they have said that they believe the answer is equally likely to be above $\mu$. That is, they believe it is equally likely for a message from True News to say ``Greater than $\mu$'' and ``Less than $\mu$,'' and equally likely for a message from Fake News to say ``Greater than $\mu$'' and ``Less than $\mu$.'' Therefore, seeing a message that says ``Greater than $\mu$'' will reveal nothing to a Bayesian about the likelihood that the message comes from the True or Fake source. However, a motivated reasoner who is motivated to believe that the answer is greater than $\mu$ will believe that the ``Greater than $\mu$'' message is more likely than the ``Less than $\mu$'' message to come from True News.

I test for valence-driven motivated reasoning by comparing how participants assess the veracity of good news and bad news. I frame the main hypothesis as an if-then statement:
\begin{hypothesis}
    If participants are susceptible to motivated reasoning and valence positively affects motivated beliefs, then they will give higher assessments to good news than to bad news, and vice versa if valence negatively affects motivated beliefs. 
    
    Conversely, if they are susceptible to motivated reasoning and valence does not affect motivated beliefs, then they will give the same assessments to good news and bad news.
\end{hypothesis}
In the experiment, we will see that participants give similar assessments to good news and bad news. By contraposition to the first statement, this could indicate either that either valence does not affect motivated beliefs, or that participants are not susceptible to the bias at all. To rule out the second explanation, we consider what happens in other domains:
\begin{hypothesis}
    If participants are not susceptible to motivated reasoning, then they will not give higher assessments to news that supports other domains of motivated reasoning (such as politics and performance). 
\end{hypothesis}
By contraposition, if people assess pro-party and pro-performance news to be more truthful than anti-party and anti-performance news, then they are susceptible to motivated reasoning in general, and thus we can conclude that the null effect is because there are not directional motivated beliefs about valence.

\subsection{Details}
\label{experiments}



The timing and payment work as follows: Participants first see an introduction page for consent, then a demographics page, and then the instructions and point system for question pages. Next, participants see the instructions and point system for news pages. Participants see news pages immediately after their corresponding Question page: Question 1, News 1, Question 2, News 2, .... At the end of the experiment, they see their performance, correct answers, and bonus. They earn a show-up fee of \$3 and either receive an additional \$10 bonus or no additional bonus. In each round of the experiment participants earn between 0-100 ``points'' based on their previous answers. These points correspond to the probability that the participant wins the bonus: A score of $x$ points corresponds to an $x/10$ percent chance of winning the bonus.\footnote{This earnings system is similar to the most broadly incentive-compatible one from \textcite{ACH18} in which participants are paid randomly for one round. I use my procedure instead in order to allow for a clearer measure of ``performance'' that is used as a question in the experiment. I do not need to assume risk neutrality in order for the experiment to be incentive compatible, but I do need to assume linearity in probabilities.}

The experiment was conducted on Amazon's Mechanical Turk (MTurk) platform. MTurk has become a popular way to run economic experiments (\cite{HRZ11}), and \textcite{LFD16} find that participants generally tend be more diverse than students in university laboratories on dimensions like age and politics. The experiment was coded using oTree, an open-source software based on the Django web application framework (\cite{CSW16}). 

The experiment was run in two waves. Wave 1 was conducted in July 2019, and asked about one non-political question with valence (about leukemia survival rates) as well as six political questions that are also classified as having valence.\footnote{A previous version of this paper removed these observations in the analysis. However, helpful suggestions have led to their re-inclusion.} Wave 1 additionally included political and performance questions that were not coded as having valence and are thus not in these analyses. It had 13 rounds in total. For 34 percent of participants in Wave 1, all rounds are used. For the remaining 66 percent of participants, only the first three rounds are used. The remaining rounds come after a treatment that is not studied in this paper. Results are robust to the inclusion of treated participants. Questions are in random order, except for round 13, which asks about performance on the previous questions.

Wave 2 was conducted in October 2019 and asked about the other four valence questions. No additional treatments were run. Both waves were offered to MTurk workers currently living in the United States who had not previously participated in my experiments. 1,050 participants from Wave 1 and 508 participants from Wave 2 passed simple attention and comprehension checks.\footnote{In order to pass these checks, participants needed to correctly answer the comprehension check question in \Cref{comprehension-question} by giving a correct answer, bounds, and news assessment. In addition, many questions had clear maximum and minimum possible answers; participants were dropped if any of their answers did not lie within these bounds.} As shown in the Online Appendix, results are robust to the inclusion of these participants.

There are a total of 5,731 initial guesses: 3,699 guesses from Wave 1 and 2,032 guesses from Wave 2. 39 (0.7 percent) of these guesses are exactly correct.\footnote{The low share of correct guesses suggests, reassuringly, that participants were not typically looking up the correct answers.} There are therefore a total of 5,692 news assessments from 1,521 participants. The Online Appendix confirms that treatments were balanced across demographic measures.

\section{Results}\label{results-main}

\subsection{Raw Data}

This subsection shows that the raw data supports cannot reject the hypothesis of no motivated reasoning towards good news or bad news. The following subsection shows the relevant regressions. 

The mean assessment of good news is 57.5 percent (s.e.\ 0.5 pp) and the mean assessment of bad news is 57.8 percent (s.e.\ 0.5 pp).\footnote{Assessments that are greater than 50 percent are typical in papers using this design (e.g. \cite{T-WP}), as participants seem to either believe that the true likelihood is greater than 50 percent, or infer credibility from messages in general. The average assessment on a neutral question is similar.} All standard errors are clustered at the individual level. The difference between these assessments is -0.3 percentage points; this point estimate is statistically insignificantly different from zero ($p=0.625$). As discussed later, the estimate is virtually unchanged in regression specifications that use within-participant tests, and the null estimate is precise. 

\begin{figure}[!htb]
\caption{CDFs of News Assessments}
\label{raw-cdfs}
\vspace{-5mm}
\begin{center}
\includegraphics[width = .7\textwidth]{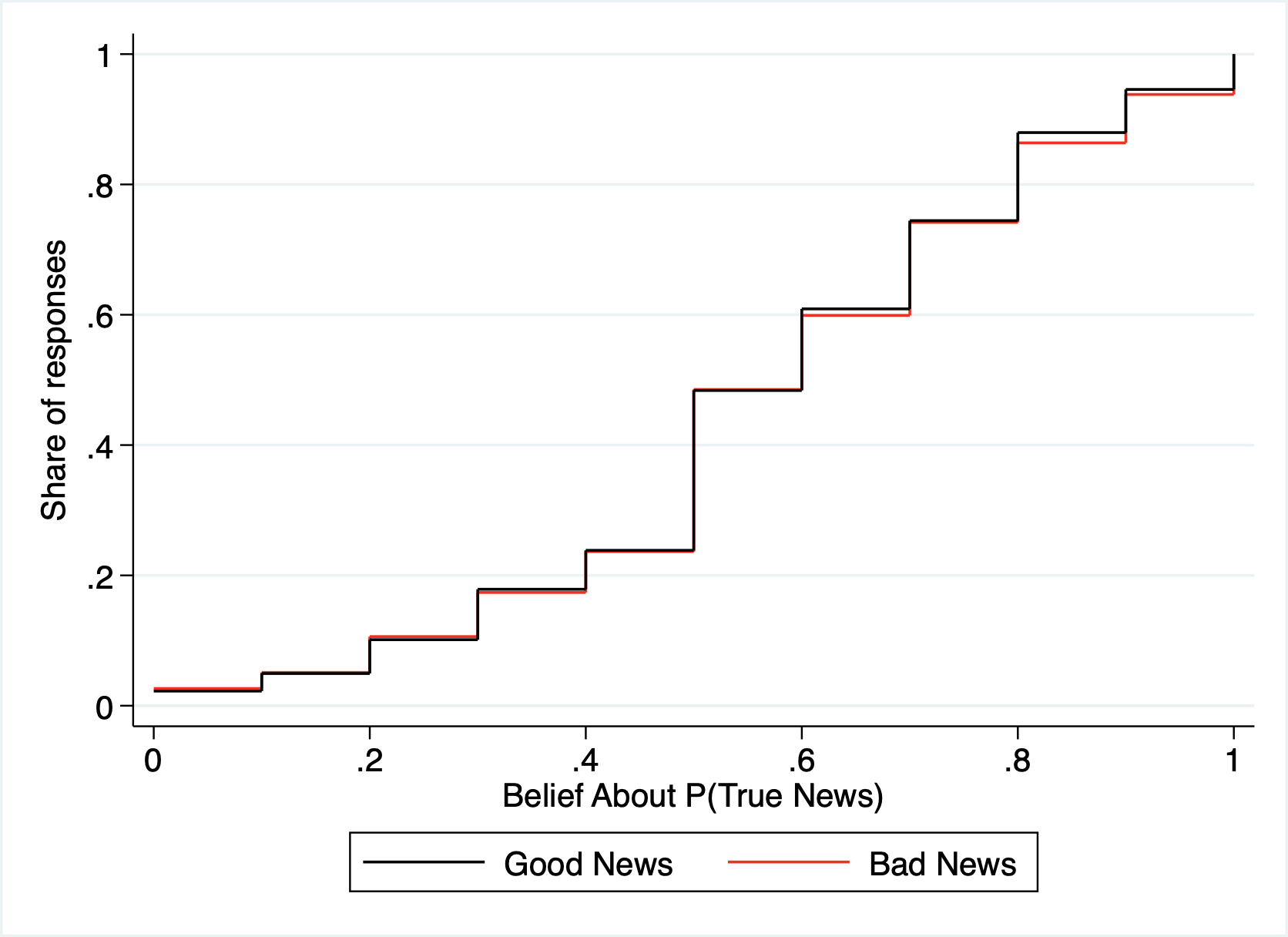}
\end{center}
\vspace{-7mm}
\begin{center}
\includegraphics[width = .7\textwidth]{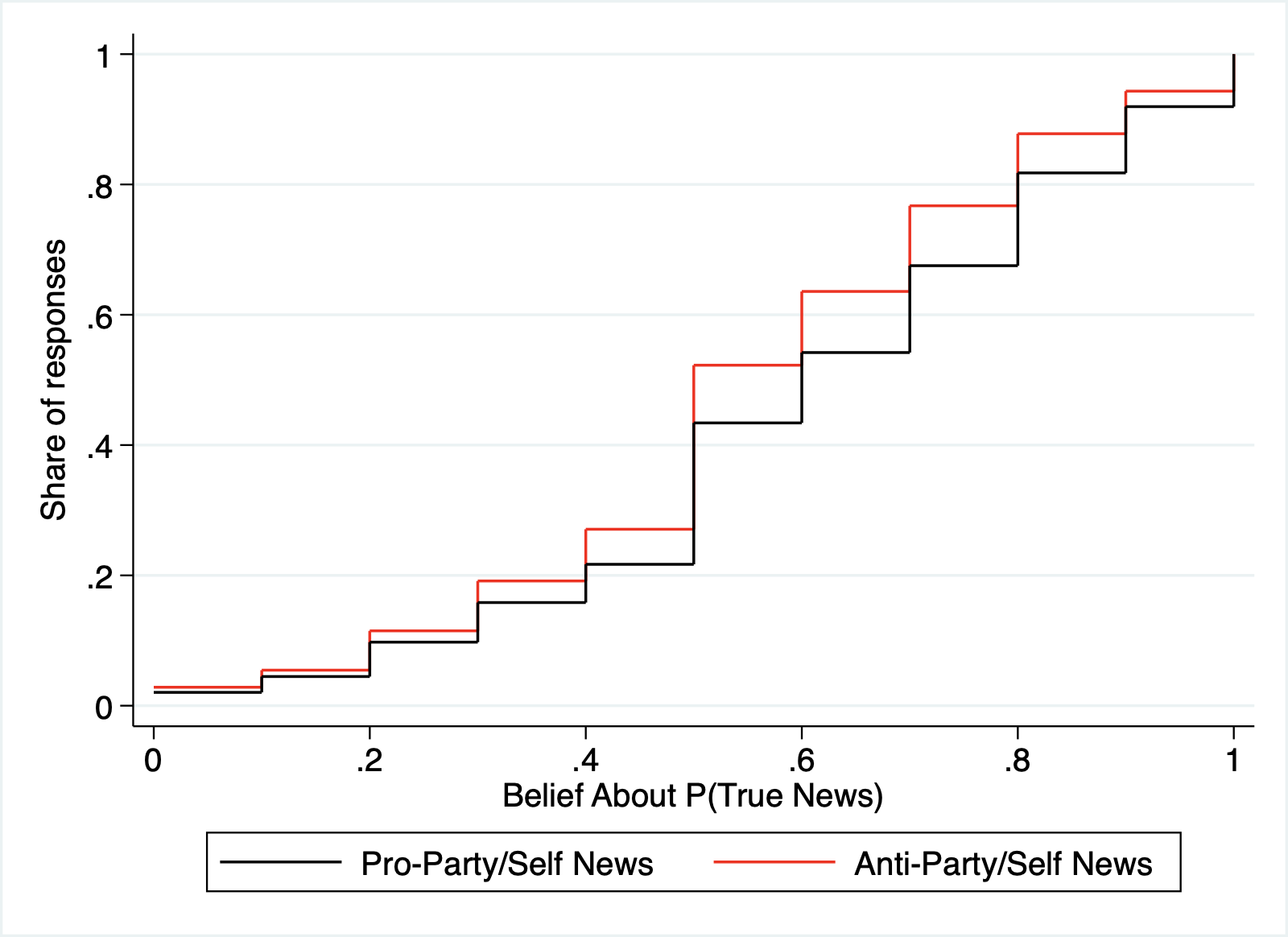}
\end{center}
\vspace{-8mm}
\begin{threeparttable}
\begin{tablenotes}
\begin{scriptsize}
\item \textbf{Notes:} This figure plots empirical CDFs for news assessments about good news and bad news (top panel) and about pro-party/self news and anti-party/self news (bottom panel). It shows no difference between good and bad news assessments, but significant differences between other assessments. Both waves included. Good news, bad news, pro-party/self, and anti-party/self news are described in Table 2.
\end{scriptsize}
\vspace{-8mm}
\end{tablenotes}
\end{threeparttable}
\end{figure}

The result that the average assessments of good and bad news are similar is not because there are different distributions of responses. The top panel of \Cref{raw-cdfs} shows the empirical CDFs of assessments for good and bad news. The lines lie on top of each other, indicating that the distributions of assessments for good and bad news are nearly-identical. This can be seen more formally using the Kolmogorov-Smirnov test, which shows that the null of the CDFs being different cannot be ruled out ($p=0.873$). The bottom panel of \Cref{raw-cdfs} shows the empirical distributions of assessments for news about party and performance; here, the assessments of pro-party and performance news first-order stochastically dominate those of anti-party/performance, and the CDFs are significantly different (Kolmogorov-Smirnov test: $p<0.001$).

Initial beliefs are not systematically biased towards good news; in fact, a slight majority (54.6 percent) of initial guesses are pessimistic. By contrast, 60.1 percent of initial guesses are more pro-party than the truth, and 58.4 percent of participants overstate their performance. Importantly, optimism in initial beliefs does not predict optimism in news assessments. On non-political topics,\footnote{We exclude political topics here, since initial beliefs are correlated with news assessments because both are correlated with participants' ideologies.} for participants with optimistic (pessimistic) priors, the mean assessment of good news is 0.7 pp lower (s.e.\ 1.7 pp) (0.8 pp lower; s.e.\ 1.3 pp). These differences are small, statistically indistinguishable from zero, and indistinguishable from each other.

By contrast, politically-motivated assessments are driven by participants with overly pro-party initial beliefs. For participants with pro-party (anti-party) priors, the mean assessment of pro-party news is 7.5 pp higher (s.e.\ 1.0 pp) (0.4 pp higher; s.e.\ 1.2 pp). Similarly, most of the performance effect is driven by participants who are optimistic about their performance. For participants with optimistic (pessimistic) priors about their performance, the mean assessment of pro-performance news is 11.8 pp higher (s.e.\ 3.3 pp) (3.1 pp higher; s.e.\ 4.2 pp). 

Taken together, these results indicate that there is no evidence for valence-driven motivated reasoning, and that optimistic priors are not predictive of motivated reasoning absent political and performance factors.

\subsection{Regression Specifications}

To formalize the analyses above, \Cref{regression-assessment} presents between-person and within-person regression specifications. Column 1 uses a between-person specification, looking at assessments $a$ for participant $i$, question topic $q$, and round $r$ when news is either good or bad:
\begin{equation*}
a_{iqr} = \alpha + \beta \cdot 1(\text{Good News})_{iqr} + \gamma^{\text{controls}} z_i + \delta FE_q + \zeta FE_r + \epsilon_{iqr}.
\end{equation*}
$z_i$ is a vector of controls: age, political party, race, gender, log(income), years of education, and religious group indicator. Fixed effects are included at the question and round level. 

Column 2 replaces the vector of controls with individual-level fixed effects, $FE_i$:
\begin{equation*}
a_{iqr} = \alpha + \beta \cdot 1(\text{Good News})_{iqr} + \gamma^{\text{FE}} FE_i + \delta FE_q + \zeta FE_r + \epsilon_{iqr}.
\end{equation*}
The within-person specification is natural since this represents the level of treatment randomization. One drawback is that some participants randomly happen to not see both good and bad news, and 4 percent of singleton observations are dropped. Reassuringly, the specification has minimal impact on estimates: In both versions, $\beta$ is precisely estimated to be close to zero. 

Columns 3 and 4 consider both the effect of valence and the effect of politics, restricting the sample to participants in Wave 1 who saw both types of questions. Column 3 uses controls, and column 4 uses participant fixed effects. The treatment effect of good news is still at zero, while the treatment effect of pro-party news is positive and statistically significant. The coefficient on pro-party news is statistically significantly larger than the coefficient on good news ($p<0.001$). The performance question cannot be directly compared, as it is not coded as having valence, but the treatment effect on it is even larger at 8.5 pp (s.e.\ 2.7 pp).

\begin{footnotesize}
\begin{center}
\begin{threeparttable}[!htb]
\caption{The Effect of Good News on Motivated Reasoning}
\begin{tabular}{l*{8}{c}}
\hline\hline
                    &\multicolumn{1}{c}{(1)}&\multicolumn{1}{c}{(2)}&\multicolumn{1}{c}{(3)}&\multicolumn{1}{c}{(4)}&\multicolumn{1}{c}{(5)}&\multicolumn{1}{c}{(6)}&\multicolumn{1}{c}{(7)}&\multicolumn{1}{c}{(8)}\\
                    &\multicolumn{4}{c}{Dep Var: News Assessments}&\multicolumn{4}{c}{Dep Var: Changing Guesses}\\
\hline
Good News           &   -0.002&   -0.002&   -0.000&    0.002&   -0.003&    0.005&    0.021&    0.021\\
                    &  (0.006)&  (0.007)&  (0.008)&  (0.009)&  (0.016)&  (0.017)&  (0.024)&  (0.024)\\
Pro-Party News      &         &         &    0.045&    0.054&         &         &    0.043&    0.043\\
                    &         &         &  (0.009)&  (0.010)&         &         &  (0.024)&  (0.024)\\
Question FE         &      Yes&      Yes&      Yes&      Yes&      Yes&      Yes&      Yes&      Yes\\
Round FE            &      Yes&      Yes&      Yes&      Yes&      Yes&      Yes&      Yes&      Yes\\
Subject FE          &       No&      Yes&       No&      Yes&       No&      Yes&      Yes&      Yes\\
Participant controls &      Yes&       No&      Yes&       No&      Yes&       No&      Yes&       No\\
\hline
Observations        &     5692&     5464&     2964&     2706&     5692&     5464&     2706&     2706\\
Participants        &     1521&     1293&      916&      658&     1521&     1293&      658&      658\\
Mean                &    0.576&    0.576&    0.575&    0.575&    0.635&    0.635&    0.661&    0.661\\
\hline\hline
\multicolumn{9}{l}{\footnotesize Standard errors in parentheses}\\
\end{tabular}

\label{regression-assessment}
\begin{tablenotes}
\begin{scriptsize}
\item \textbf{Notes:} OLS, errors clustered at participant level. Both waves included. All classifications are described in Table 2. Controls: age, political party, race, gender, log(income), years of education, and membership in a religious group. Columns with Pro-Party News include only questions that have both valence and politics, i.e. rows 6-11 of Table 2. A small number of observations are dropped when Participant FE are included, as some participants only see all Good News or all Bad News by chance.
\end{scriptsize}
\end{tablenotes}
\end{threeparttable}
\vspace{5mm}
\end{center}
\end{footnotesize}

Good news does not impact motivated reasoning, as measured by news assessments, at any significant level, and modest effect sizes can be statistically ruled out. Consistent with past evidence, political (and performance) domains lead to motivated reasoning. That is, from Hypotheses 1 and 2 we can infer that participants \textit{are} susceptible to motivated reasoning, but that valence does not trigger the bias.

Results are similar if we use an alternative measure of motivated reasoning. Here, we consider how participants update their beliefs about the original question from $\mu$ to $\mu'$. I define \textit{Follow Message} as the ternary variable that takes value: 
\begin{itemize}
\item 1 if the participant sees a ``Greater than $\mu$'' message and $\mu' > \mu$, or if they see a ``Less than $\mu$'' message and $\mu' < \mu$;
\item 0 if $\mu' = \mu$; and
\item -1 if the participant sees a ``Greater than $\mu$'' message and $\mu' < \mu$, or if they see a ``Less than $\mu$'' message and $\mu' > \mu$.
\end{itemize}
The last four columns of \Cref{regression-assessment} use the same specifications as \Cref{regression-assessment}, but with Follow Message as the dependent variable, reaching similar conclusions.\footnote{Effects of pro-party news point in the same direction as before, but this estimate has less precision. If all political questions, and not just those that are classified as good/bad news, are considered, the coefficient for pro-party news is 0.038 (s.e.\ 0.018), with $p=0.036$.}

\subsection{Heterogeneity}

The results above have shown that the \textit{average} level of valence-driven motivated reasoning is close to zero. This may be because few people engage in motivated reasoning about these topics, or it may be because there is ample motivated reasoning, but heterogeneity across people that happens to have mean zero. Here, I argue that results are most consistent with the former interpretation, finding no systematic evidence that certain types of people are ``optimistically'' motivated while others are ``pessimistically'' motivated.

I consider three forms of heterogeneity: prior beliefs, participant demographics, and issue type. \Cref{heterogeneity-positivity} plots the treatment effect on good news interacted with each of these sets of variables. It also provides benchmarks for no motivated reasoning, the treatment effect for pro-party news, and the treatment effect for pro-performance news.

\begin{footnotesize}
\centering
\begin{figure}[!htb]
\caption{Heterogeneity in Motivated Reasoning Towards Good News}
\begin{center} \includegraphics[width = .9\textwidth]{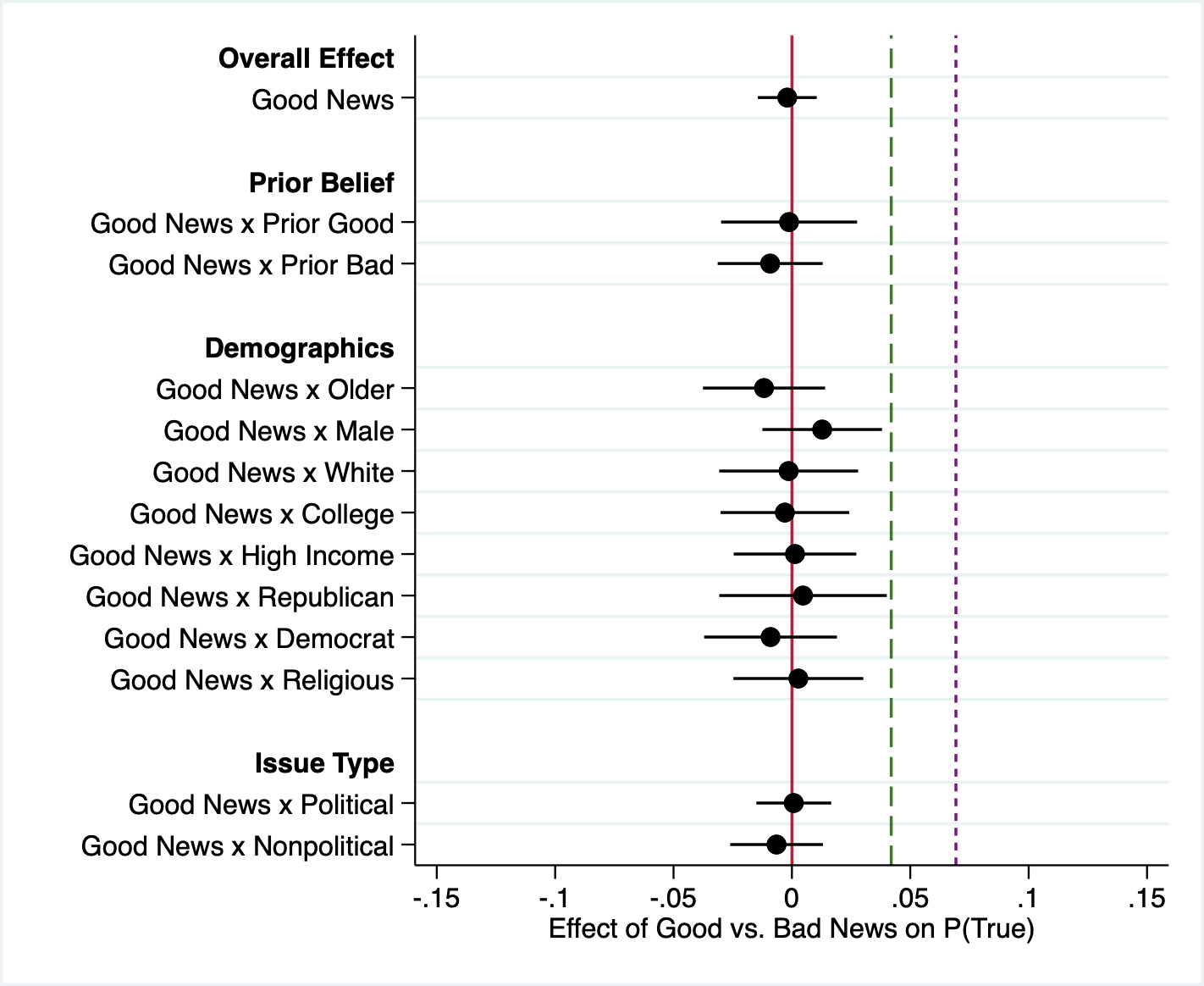} \end{center}
\label{heterogeneity-positivity}
\begin{threeparttable}[htb!]
\begin{tablenotes}
\begin{scriptsize}
\item \vspace{-8mm}\textbf{Notes:} OLS regression coefficients, errors clustered at participant level. FE included for round number and topic. Only good/bad news observations, as described in Table 2. Solid red line: Bayesian benchmark of zero treatment effect. Long-dashed green line: treatment effect of pro-party news. Short-dashed purple line: treatment effect of pro-self news. Religious group: participant affiliates with any religion. Age and income cutoffs are at the median. Prior Good/Bad: initial guess was biased in the good/bad direction. Error bars correspond to 95 percent confidence intervals. 
\end{scriptsize}
\end{tablenotes}
\end{threeparttable}
\end{figure}
\end{footnotesize}

First, one prediction of motivated reasoning is that current beliefs reflect motivated beliefs. For instance, on political topics, participants who currently hold beliefs that are biased towards Republican stances assess news that says their answers should be further in the pro-Republican direction as more likely to be true, compared to news that says their answers should be more moderate, and likewise for Democratic beliefs.\footnote{See \textcite{T-WP} for a more in-depth discussion. In performance domains, \textcite{ER11} shows that priors are related to motivated reasoning, and \textcite{T21} shows that gender differences in overconfidence and motivated reasoning are correlated.} However, that heterogeneity is not present on valence questions. As shown by the second set of coefficients in \Cref{heterogeneity-positivity}, participants whose beliefs are overly optimistic and participants whose beliefs are overly pessimistic both have null treatment effects to good news versus bad news.

Second, motivated reasoning about good news may vary across demographic lines. For instance, participants who have higher income may be more optimistic, and gender and politics often play roles in other motivated-reasoning domains. However, the third set of coefficients in \Cref{heterogeneity-positivity} show that none of these demographics (nor race, education, age, or religiosity) directionally affect the estimate in any meaningful (or statistically significant) way. In none of these categories do the treatment-effect error bars overlap with the overall treatment effect in the political or performance domains. 

Third, it is possible that issues evoke motivated reasoning differently when they are political versus non-political. For instance, perhaps bad news for political issues may be more attractive because it mimics what news emphasizes, while good news is more attractive for non-political issues. However, there is no evidence for this discrepancy: the last set of coefficients in \Cref{heterogeneity-positivity} finds no effect of issue type on treatment effects. 

The data are not precise enough to clearly argue in favor of or against heterogeneity between topics. When we account for multiple hypothesis testing, the treatment effect on all but one topic is statistically insignificant at the $p=0.050$ threshold.\footnote{$p$-values use the \textcite{WY93} approach from \textcite{JMR19}.} The one exception is global poverty, where there is statistically significant evidence for motivated reasoning in the \textit{pessimistic} direction.\footnote{One possibility is that this question is not solely about valence, and evokes social-comparison or political concerns.} Dropping this question only slightly changes treatment effects (0.7 pp; s.e.\ 0.7 pp). 

In in the Online Appendix, I discuss three additional robustness checks. First, while my main specification uses linear probabilities, regressions in these settings are sometimes written in logit form (\cite{G80}). I show that results are similar if logit news assessments are used as the dependent variable instead. Second, including participants who failed comprehension checks in the analyses also does not substantively change the results. Lastly, one may be concerned that the experimental design lends itself to strategic misreporting of beliefs because participants misreport their median in order to make news assessment questions easier. While a theoretical concern, the data show nearly-identical treatment effects in Round 1 of the experiment (in which participants did not yet know they would be given a news assessment page) as compared to later rounds (in which they did know).

\section{Discussion and Survey Results}
\label{discussion}

Results from the experiment indicate that there is little impact of good news or bad news on motivated reasoning, absent self-relevant or functional factors, and that the null effect is robust and precise. We now consider two explanations for these results. (1) People do not internalize positive and negative emotions when they receive good news or bad news about these issues. (2) Motivated beliefs are systematically different from belief-based utility, leading people to process information differently in valence domains versus political and performance domains. 

To disentangle these two explanations, I ran two follow-up surveys among a new group of participants --- drawn from different MTurk samples --- in January 2020. I recruited 303 participants in Survey 1 and 167 in Survey 2.\footnote{Participants were required to not have taken the original experiment. I additionally excluded 16 participants in Survey 1 and 5 participants in Survey 2 who failed an attention check.} Further details and screenshots are in the Online Appendix. Survey evidence is more consistent with explanation (2). That is, the results indicate that participants expect motivated beliefs to be systematically different from belief-based utility. 

In Survey 1, participants are asked to predict the direction of others' motivated reasoning about good news, politics, and performance (in random order), and are given example topics for each.\footnote{Participants do not do the main experiment themselves, in order to avoid unintentional treatment effects from the experiment on beliefs.} For good news, they were asked whether they thought that most people motivatedly reasoned in the direction of believing that the world was a better place for others, most people motivatedly reasoned in the direction of believing the world was a worse place for others, or about the same.

\begin{footnotesize}
\centering
\begin{figure}[!htb]
\caption{Survey Evidence: Motivated Reasoning and Belief-Based Happiness}

\begin{subfigure}[b]{\textwidth}
\caption{Survey 1: Does Good News Lead to Motivated Reasoning?}
\vspace{-5mm}
\begin{center}
\includegraphics[width = .68\textwidth]{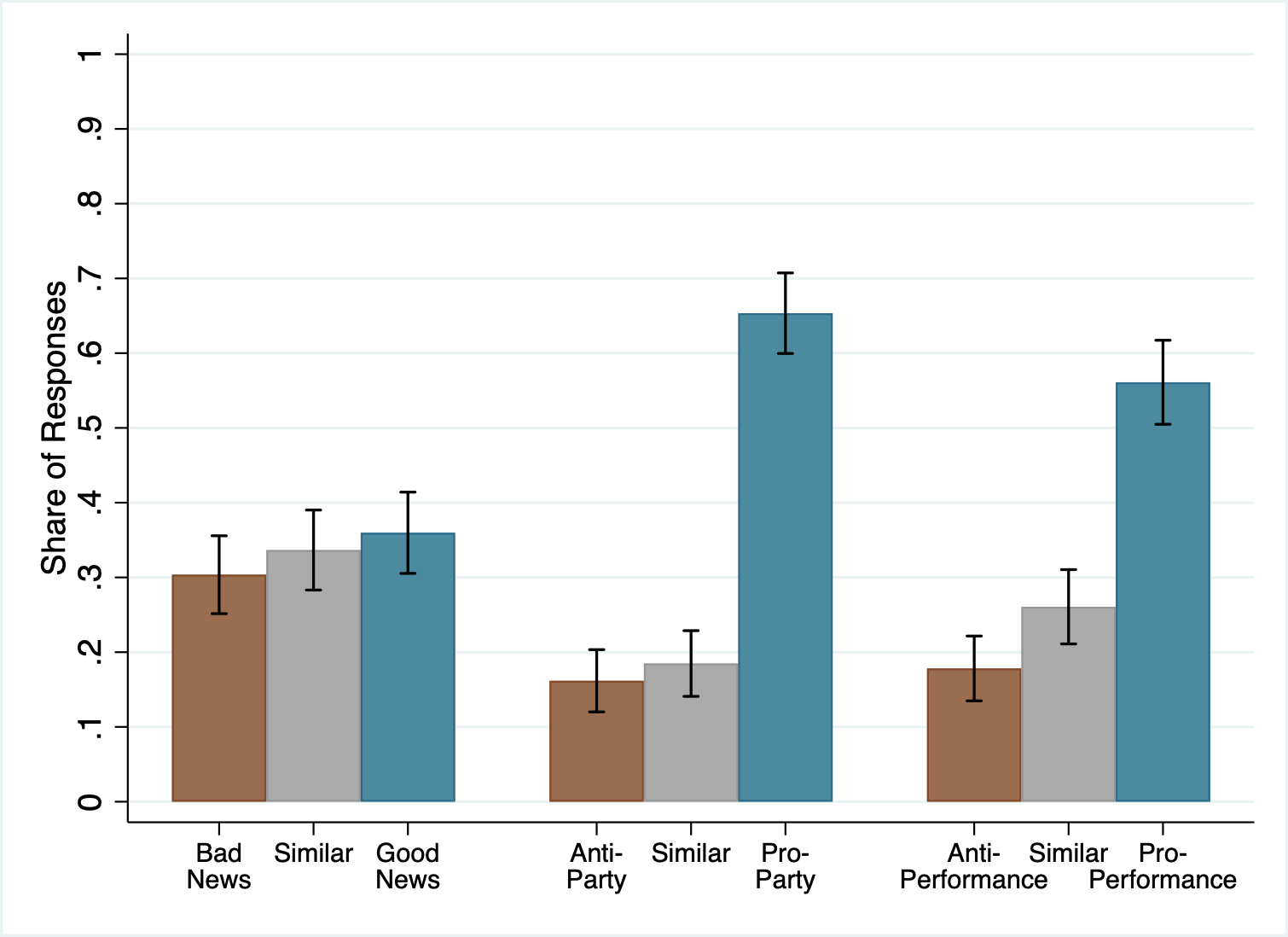}
\end{center}
\label{positivity-survey}
\end{subfigure}

\begin{subfigure}[b]{\textwidth}
\caption{Survey 2: Does Good News Lead to Happiness?}
\vspace{-5mm}
\begin{center}
\includegraphics[width = .68\textwidth]{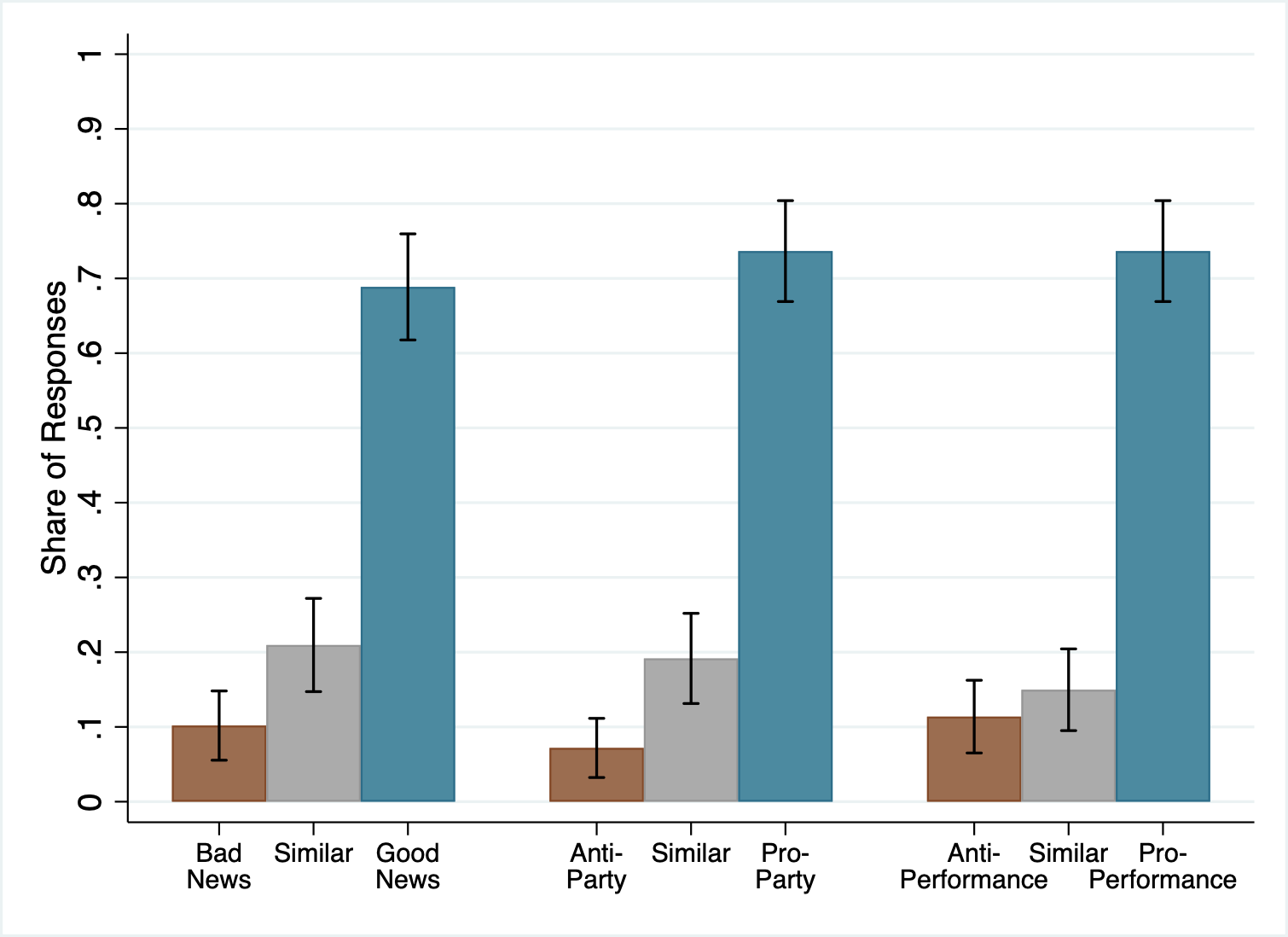}
\end{center}
\label{positivity-survey-happiness}
\end{subfigure}
\begin{threeparttable}
\begin{tablenotes}
\vspace{-9mm}
\begin{scriptsize}
\item \textbf{Notes:} The y-axis is the share of respondents who stated that they expect most people (a) to have motivatedly reasoned, or (b) to be happier, given news in one direction, the other direction, or a similar amount in both directions. 0.2 percent of questions are left unanswered, and are coded as ``Similar.'' Error bars correspond to 95 percent confidence intervals. 
\end{scriptsize}
\end{tablenotes}
\end{threeparttable}
\vspace{-10mm}
\end{figure}
\end{footnotesize}

Results from Survey 1 are shown in \Cref{positivity-survey}. 65 percent of participants expected motivated-reasoning distortions in the pro-party direction, versus only 16 percent who expected anti-party distortions. Similarly, 56 percent expect pro-performance distortions, and only 18 percent expect anti-performance distortions. However, participants were similarly likely to predict distortions towards good news (36 percent) and bad news (30 percent); this difference is not statistically significant ($p=0.231$). They were also more likely to predict that there would not be directional distortions about valence, as compared to party and performance. These differences are statistically significant ($p<0.005$ each). This suggests participants correctly anticipate that there is systematic directional motivated reasoning about politics and performance, but not about valence.

In Survey 2, participants were given the same categories and examples, but were now asked to predict whether such beliefs would lead people to be \textit{happier}.\footnote{While the emotion of happiness has differences from the economic notion of utility, I used ``happiness' for ease of participant understanding.} Questions were identical to the page from Survey 1, but replaced the phrase ``Do you think people tended to motivatedly reason in favor of believing...?'' with ``Do you think people were happier when they received information that supported the belief...?''.

Results from Survey 2 are shown in \Cref{positivity-survey-happiness}. Unlike Survey 1, a clear majority (69 percent, s.e.\ 3.6 pp) of respondents expected good news to make people happier, and very few (10 percent, s.e.\ 2.3 pp) expected bad news to increase happiness. The effect is similar for political and performance questions; none of the differences across topics are statistically significant ($p>0.05$).

These results indicate that most people expect good news to increase happiness, just as good news about politics or performance. However, while most people expect motivated reasoning to affect inference, they do not expect the bias to be domain general. Rather, they believe that happiness is insufficient to induce the bias. The gap may be explained by people feeling happier from believing in --- but not motivatedly reasoning about --- others' well-being. 

\section{Conclusion}
\label{conclusion}

This paper has shown that people do not engage in motivated reasoning in the direction of ``good'' states of the world when other factors like functional and ego concerns are not at play. When given good news versus bad news, absent these factors, people do not update their beliefs differently. There is also no evidence for optimistic or pessimistic ``types'' in the population. These results are in stark contrast to the robust evidence for motivated reasoning in political and performance domains. Survey results additionally showed that people do not expect that good news will induce motivated reasoning, while they do think that good news leads to greater happiness. That is, there are many reasons why people form persistently-inaccurate beliefs, but maximizing happiness does not seem to be a crucial factor.

These results suggest many potential directions for future work. For instance, how does belief-based utility, absent social and ego-relevant factors, affect choice behavior? The literature documents empirical patterns on preferences for information (e.g. \cite{OSD13}; \cite{GHL17}; \cite{GT16}), and further work could unpack whether this is primarily driven by good and bad news in general, or by functional and ego-relevant factors specifically. Further work could also extend the approach of \textcite{L00} to understand how emotions affect information processing, and whether other patterns of belief updating have emotional underpinnings. Another fruitful possibility is to extend the experimental design to study how people process real news articles as a function of the articles' valence; documenting information-processing patterns outside the lab would contribute to our understanding of the effect of media on beliefs.

\vspace{10mm}

\printbibliography

\newpage 

\appendix

\section*{\centering
\huge{Supplementary Appendix}}
\section{Experiment Topics}

\begin{landscape}

\renewcommand\arraystretch{1.5}
\begin{table}
\begin{center}
\begin{tabular}{l l l l l}  
\toprule
\textbf{Topic}  & \textbf{Positive} & \textbf{Negative} & \textbf{Democratic} & \textbf{Republican} \\
\midrule
Infant mortality & Low / Decreasing & High / Increasing & N/A & N/A \\
Others' reported happiness & High / Increasing & Low / Decreasing & N/A & N/A \\
Child leukemia survival rate & High / Increasing & Low / Decreasing & N/A & N/A \\
Global poverty rate & Low / Decreasing & High / Increasing & N/A & N/A \\
Deaths in armed conflicts & Low / Decreasing & High / Increasing & N/A & N/A \\
Wage growth & High & Low & Higher under Obama & Higher under Trump \\
Students' math GPA & High & Low & Higher for HS girls & Higher for HS boys \\
Job callback rates & High & Low & More racial discrimination & Less racial discrimination \\
Gun deaths & Low & High & Lower after gun laws & Higher after gun laws \\
Violent crime & Low & High & Lower among refugees & Higher among refugees \\
Murder rates & Low & High & Decreased under Obama & Increased under Obama \\
Performance & N/A & N/A & High & High \\
\hyperref[latitude-question]{Latitude of US} & Neutral & Neutral & Neutral & Neutral \\
\bottomrule
\end{tabular}
\caption{The list of topics and hypothesized directional motives in the experiment. Exact question wordings are in the Online Appendix.}
\label{topics-positivity}
\end{center}
\end{table}

\end{landscape}

\renewcommand\arraystretch{1}
\vspace{-5mm}

\clearpage

\section*{\centering
\huge{Online Appendices}}
\setcounter{page}{1}

\section{Additional Tables and Figures}



\subsection{Balance Table}
\vspace{3mm}

\begin{center}
\def\sym#1{\ifmmode^{#1}\else\(^{#1}\)\fi}
\begin{tabular}{p{2.5cm}*{4}{c}}
\hline\hline
&\multicolumn{1}{c}{Bad News}&\multicolumn{1}{c}{Good News}&\multicolumn{1}{c}{Bad vs. Good}&\multicolumn{1}{p{2.75cm}}{\centering p-value}\\
\hline
Male  & 0.542 & 0.527 & 0.015 & 0.255 \\   & (0.009) & (0.009) & (0.013) \\Age  & 35.573 & 35.639 & -0.066 & 0.817 \\   & (0.205) & (0.201) & (0.287) \\Education  & 14.833 & 14.801 & 0.032 & 0.531 \\   & (0.036) & (0.036) & (0.051) \\Log(Income)  & 10.797 & 10.823 & -0.026 & 0.221 \\   & (0.015) & (0.015) & (0.021) \\Democrat  & 0.467 & 0.460 & 0.007 & 0.598 \\   & (0.009) & (0.009) & (0.013) \\Republican  & 0.196 & 0.193 & 0.003 & 0.787 \\   & (0.008) & (0.007) & (0.010) \\White  & 0.740 & 0.738 & 0.002 & 0.888 \\   & (0.008) & (0.008) & (0.012) \\Black  & 0.086 & 0.079 & 0.007 & 0.311 \\   & (0.005) & (0.005) & (0.007) \\Latino  & 0.059 & 0.061 & -0.002 & 0.747 \\   & (0.004) & (0.004) & (0.006) \\Asian  & 0.080 & 0.082 & -0.003 & 0.725 \\   & (0.005) & (0.005) & (0.007) \\Religious  & 0.484 & 0.483 & 0.001 & 0.954 \\   & (0.010) & (0.009) & (0.013) \\\hline
\(N\)  & 2761 & 2931 & 5692  & \\
\hline
\hline
\end{tabular}
\end{center}
\label{balance-table}

\begin{small} \textbf{Notes:} Standard errors in parentheses. Education is in years. Religious is 1 if participant is in any religious group. Both waves included. Only good/bad news observations, as described in Table 2. See the Study Materials for exact demographic questions. \end{small}
\clearpage

\subsection{Robustness}

\begin{footnotesize}
\begin{center}
\begin{threeparttable}[!htb]
\caption{The Effect of Good News on Trust in News: Logit Specification}
\begin{tabular}{l*{4}{c}}
\hline\hline
                    &\multicolumn{1}{c}{(1)}&\multicolumn{1}{c}{(2)}&\multicolumn{1}{c}{(3)}&\multicolumn{1}{c}{(4)}\\
                    &\multicolumn{4}{c}{Dep Var: Logit(News Assessment)}\\
\hline
Good News           &   -0.013&   -0.009&   -0.005&    0.007\\
                    &  (0.033)&  (0.034)&  (0.043)&  (0.044)\\
Pro-Party News      &         &         &    0.239&    0.291\\
                    &         &         &  (0.047)&  (0.052)\\
Question FE         &      Yes&      Yes&      Yes&      Yes\\
Round FE            &      Yes&      Yes&      Yes&      Yes\\
Subject FE          &       No&      Yes&       No&      Yes\\
Participant controls \hspace{15mm} &      Yes&       No&      Yes&       No\\
\hline
Observations        &     5692&     5464&     2964&     2706\\
Participants        &     1521&     1293&      916&      658\\
Mean                &    0.380&    0.380&    0.379&    0.379\\
\hline\hline
\multicolumn{5}{l}{\footnotesize Standard errors in parentheses}\\
\end{tabular}

\label{regression-assessment-logit}
\begin{tablenotes}
\begin{scriptsize}
\item \textbf{Notes:} This figure replicates columns 1-4 of \Cref{regression-assessment} but uses logit probability assessments on the LHS. Logit assessments are set to $\log(p)-\log(1-p)$ if participants say that P(News True) $\in [0.1, 0.9]$, set to $\log(0.95)-\log(0.05)$ if participants say that P(News True) = 1, and set to $\log(0.05)-\log(0.95)$ if participants say that P(News True) = 0. OLS, errors clustered at participant level. OLS, errors clustered at participant level. Both waves included. All classifications are described in Table 2. Controls: age, political party, race, gender, log(income), years of education, and membership in a religious group. Columns with Pro-Party News include only questions that have both valence and politics, i.e. rows 6-11 of Table 2. A small number of observations are dropped when Participant FE are included, as some participants only see all Good News or all Bad News by chance.
\end{scriptsize}
\end{tablenotes}
\end{threeparttable}
\vspace{5mm}
\end{center}
\end{footnotesize}

\clearpage

\begin{footnotesize}
\begin{center}
\begin{figure}[!htb]
\caption{Main Estimate and Heterogeneity in Motivated Reasoning: Logit Specification}
\begin{center}
\includegraphics[width = .9\textwidth]{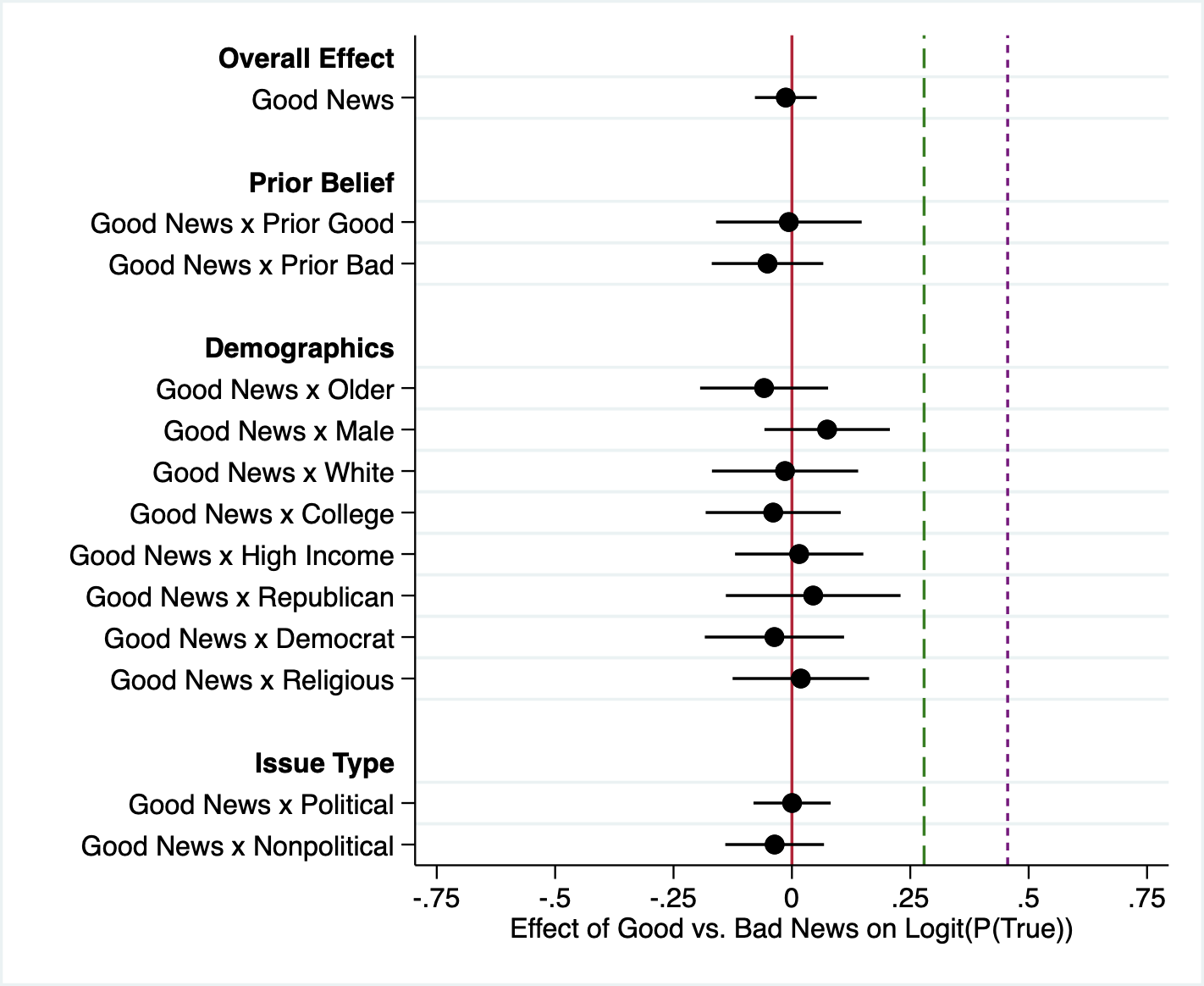}
\end{center}
\label{heterogeneity-positivity-logit}
\vspace{-5mm}
\begin{threeparttable}[htb!]
\begin{tablenotes}
\begin{scriptsize}
\item \textbf{Notes:} This figure replicates \Cref{heterogeneity-positivity} but uses logit probability assessments on the LHS. Logit assessments are set to $\log(p)-\log(1-p)$ if participants say that P(News True) $\in [0.1, 0.9]$, set to $\log(0.95)-\log(0.05)$ if participants say that P(News True) = 1, and set to $\log(0.05)-\log(0.95)$ if participants say that P(News True) = 0. OLS regression coefficients, errors clustered at participant level. FE included for round number and topic. Both waves included. Only good/bad news observations, as described in Table 2. Solid red line: Bayesian benchmark of zero treatment effect. Long-dashed green line: treatment effect of pro-party news. Short-dashed purple line: treatment effect of pro-self news. Religious group: Participant affiliates with any religion. Age and income cutoffs are at the median. Prior Good/Bad: initial guess was biased in the good/bad direction. Error bars correspond to 95 percent confidence intervals.  
\end{scriptsize}
\end{tablenotes}
\end{threeparttable}
\end{figure}
\end{center}
\end{footnotesize}

\clearpage



\begin{footnotesize}
\begin{center}
\begin{figure}[!htb]
\caption{Main Estimate and Heterogeneity in Motivated Reasoning: Including Participants Who Failed Comprehension Checks}
\begin{center}
\includegraphics[width = .9\textwidth]{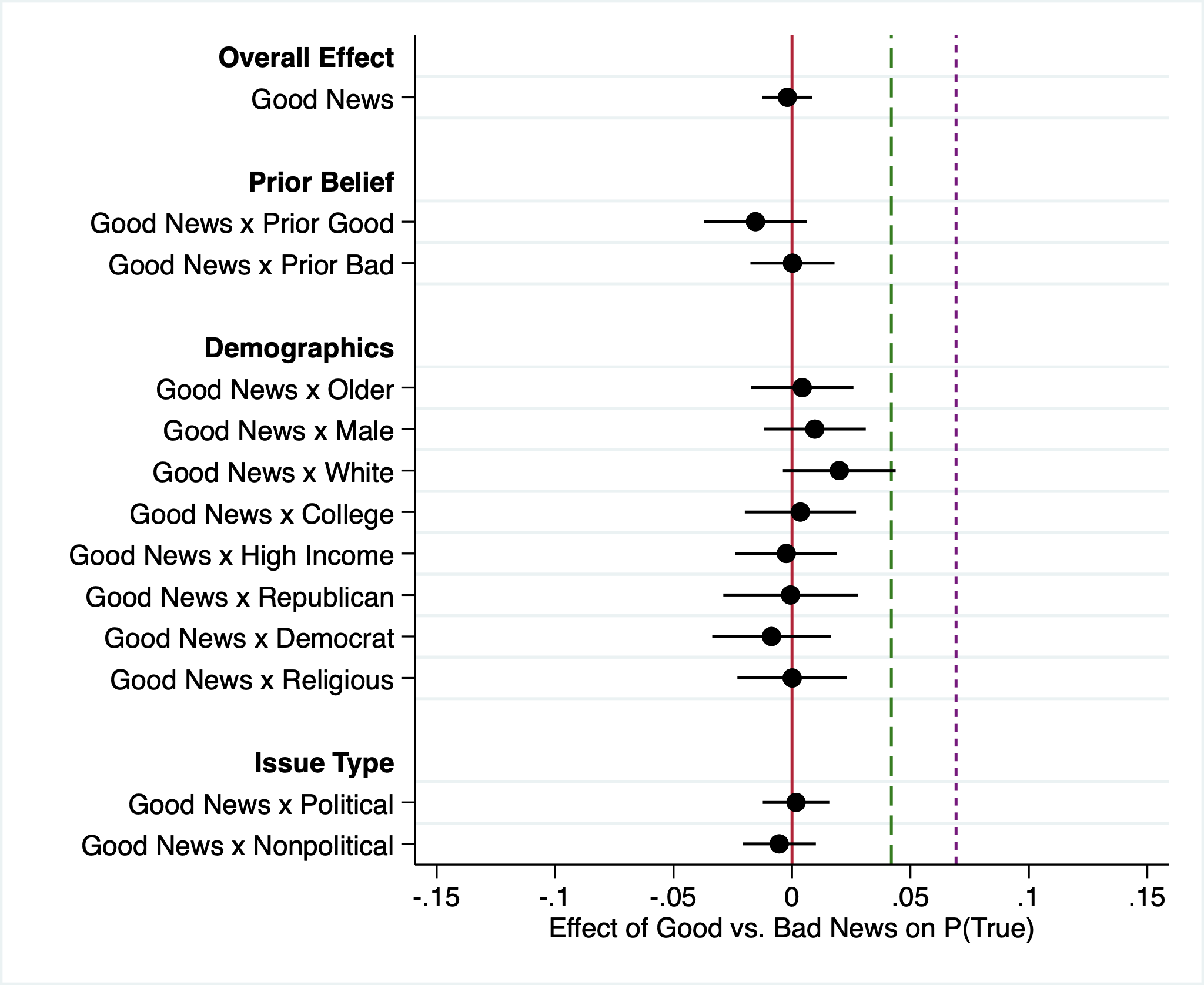}
\end{center}
\label{heterogeneity-positivity-failedcheck}
\vspace{-5mm}
\begin{threeparttable}[htb!]
\begin{tablenotes}
\begin{scriptsize}
\item \textbf{Notes:} This figure replicates \Cref{heterogeneity-positivity} but does not drop participants who failed comprehension checks. OLS regression coefficients, errors clustered at participant level. FE included for round number and topic. Both waves included. Only good/bad news observations, as described in Table 2. Solid red line: Bayesian benchmark of zero treatment effect. Long-dashed green line: treatment effect of pro-party news. Short-dashed purple line: treatment effect of pro-performance news. Religious group: Participant affiliates with any religion. Age and income cutoffs are at the median. Prior Good/Bad: initial guess was biased in the good/bad direction. Error bars correspond to 95 percent confidence intervals.  
\end{scriptsize}
\end{tablenotes}
\end{threeparttable}
\end{figure}
\end{center}
\end{footnotesize}

\clearpage

\begin{footnotesize}
\begin{center}
\begin{figure}[!htb]
\caption{Motivated Reasoning by Round}
\begin{center}
\includegraphics[width = .9\textwidth]{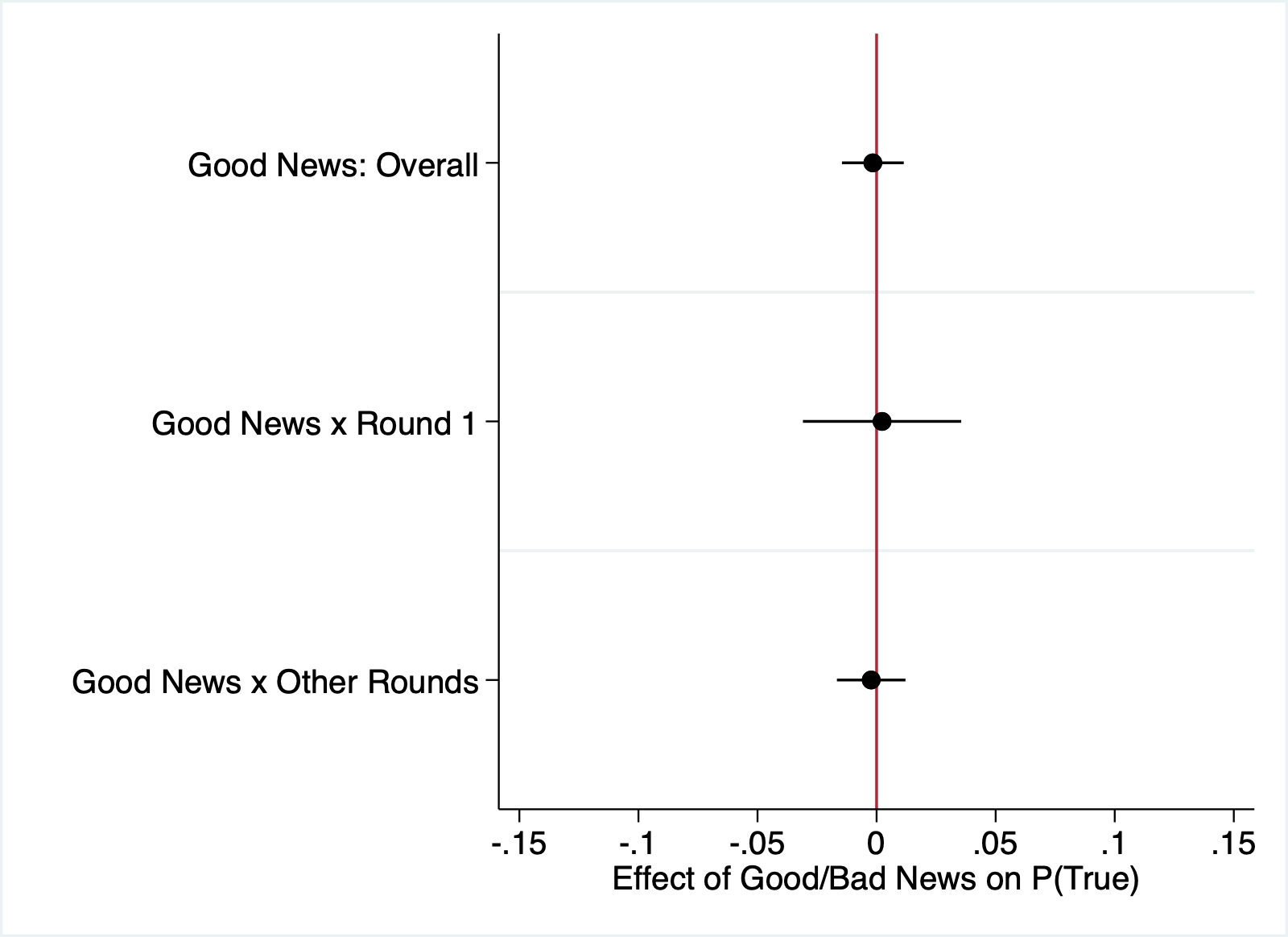}
\end{center}
\label{byround}
\vspace{-5mm}
\begin{threeparttable}[htb!]
\begin{tablenotes}
\begin{scriptsize}
\item \textbf{Notes:} OLS regression coefficients, errors clustered at participant level. Round refers to round number in experiment. FE included for round number (except for the Round 1 estimate) and topic. Both waves included. Only good/bad news observations, as described in Table 2. Solid red line: Bayesian benchmark of zero treatment effect. Error bars correspond to 95 percent confidence intervals. 
\end{scriptsize}
\end{tablenotes}
\end{threeparttable}
\end{figure}
\end{center}
\end{footnotesize}

\newpage 

\section{Study Materials: Question Wordings}
\label{question-wordings}

\subsection*{Non-Political Questions}

\subsubsection*{Cancer in Children (Wave 1)}
\label{cancer-question}
Acute Myeloid Leukemia (AML) is a devastating illness in which cancerous cells emerge in the bone marrow, invade the blood stream, and may spread to the rest of the body. Tragically, hundreds to thousands of children under the age of 15 are diagnosed with AML each year; it is one of the most common cancers among children.

Of children under the age of 15 who are diagnosed with AML, what percent survive for at least 5 years?

(Please guess between 0 and 100.)

\vspace{2mm}

\textit{Correct answer: 68.8.}

\textit{Source linked on results page: \url{https://www.lls.org/facts-and-statistics/overview/childhood-blood-cancer-facts-and-statistics}}

\subsubsection*{Infant Mortality (Wave 2)}
\label{infant-mortality-question}
The CDC provides statistics for mortality rates for infants. In 1997, there were 28.0 thousand infant deaths in the United States. 

How many thousands of infant deaths in the United States were there in 2017 (the most recent year available)? 

(If you answer X, it means you think that there were X thousand deaths.)

\vspace{2mm}

\textit{Correct answer: 22.3.}

\textit{Source linked on results page: \url{https://www.cdc.gov/nchs/data/nvsr/nvsr68/nvsr68_10-508.pdf}}

\subsubsection*{Others' Happiness (Wave 2)}
\label{happiness-question}
Many surveys ask the following question about subjective happiness:

``Please imagine a ladder with steps numbered from zero at the bottom to ten at the top. Suppose we say that the top of the ladder represents the best possible life for you and the bottom of the ladder represents the worst possible life for you.
If the top step is 10 and the bottom step is 0, on which step of the ladder do you feel you personally stand at the present time?''

In 2006, the average subjective happiness level in the United States was 7.18 out of 10.

What was the average subjective happiness level in the US in 2018?

\vspace{2mm}

\textit{Correct answer: 6.88.}

\textit{Source linked on results page: \url{https://ourworldindata.org/happiness-and-life-satisfaction}}

\subsubsection*{Global Poverty (Wave 2)}
\label{poverty-question}
Around the world, many people do not have enough money for basic necessities. The World Bank defines extreme poverty as having less than the equivalent of \$1.90 per day.

In 1990, the World Bank estimated that 1897 million people around the world were living in extreme poverty.

As of 2015 (the most recent year available), how many milllions of people around the world live in extreme poverty?

(If you answer X, it means you think that X million people live in extreme poverty.)

\vspace{2mm}

\textit{Correct answer: 731.}

\textit{Source linked on results page: \url{http://povertydata.worldbank.org/poverty/home/}}

\subsubsection*{Armed Conflict (Wave 2)}
\label{conflict-question}
The Department of Peace and Conflict Research estimates that 45.8 thousand people were killed per year in battles in the fifteen years from 1989-2003. 

How many thousands of people were killed per year in battles in the fifteen years from 2004-2018?

(If you answer X, it means you think that X thousand people were killed per year.)

\vspace{2mm}

\textit{Correct answer: 48.12}

\textit{Source linked on results page: \url{https://www.pcr.uu.se/digitalAssets/667/c_667494-l_1-k_battle-related-deaths-by-region--1989-2018.pdf}}

\vspace{-1mm}

\subsection*{Political Questions}

\subsubsection*{Wage Growth (Wave 1)}
\label{wage-question}
Participants randomly see one of the following questions:

\begin{itemize}
    \item In the last two years of Barack Obama's presidency (2015 and 2016), the median growth in Americans' wages was 3.29 percent on average.

    In the first two years of Donald Trump's presidency (2017 and 2018), what was the median growth in Americans' wages on average?

    \textit{Correct answer: 3.35 percent}

    \item In the first two years of Donald Trump's presidency (2017 and 2018), the median growth in Americans' wages was 3.35 percent on average.

    In the last two years of Barack Obama's presidency (2015 and 2016), what was the median growth in Americans' wages on average?

    \textit{Correct answer: 3.29 percent}
\end{itemize}

\vspace{-1.5mm}
\noindent \textit{Source linked on results page: \url{http://bit.ly/median-wage-growth}}

\subsubsection*{Students' Math Performance (Wave 1)}
\label{grades-question}
Participants randomly see one of the following questions:

\begin{itemize}
    \item This question asks whether high school boys and girls differ substantially in how well they do in math classes. A major testing service analyzed data on high school seniors and compared the average GPA for male and female students in various participants.

    Female students averaged a 3.15 GPA (out of 4.00) in math classes. What GPA did male students average in math classes?

    (Please guess between 0.00 and 4.00.)

    \textit{Correct answer: 3.04}

    \item This question asks whether high school boys and girls differ substantially in how well they do in math classes. A major testing service analyzed data on high school seniors and compared the average GPA for male and female students in various participants.

    Male students averaged a 3.04 GPA (out of 4.00) in math classes. What GPA did female students average in math classes?

    (Please guess between 0.00 and 4.00.)

    \textit{Correct answer: 3.15}
\end{itemize}

\vspace{-1.5mm}
\noindent \textit{Source linked on results page: \url{http://bit.ly/gender-hs-gpa}}

\subsubsection*{Job Callback Rates (Wave 1)}
\label{jobs-question}
Participants randomly see one of the following questions:

\begin{itemize}
    \item In a study, researchers sent fictitious resumes to respond to thousands of help-wanted ads in newspapers. The resumes sent had identical skills and education, but the researchers gave half of the (fake) applicants stereotypically White names such as Emily Walsh and Greg Baker, and gave the other half of the applicants stereotypically Black names such as Lakisha Washington and Jamal Jones.

    9.65 percent of the applicants with White-sounding names received a call back. What percent of the applicants with Black-sounding names received a call back?

    (Please guess between 0 and 100.)

    \textit{Correct answer: 6.45 percent}

    \item In a study, researchers sent fictitious resumes to respond to thousands of help-wanted ads in newspapers. The resumes sent had identical skills and education, but the researchers gave half of the (fake) applicants stereotypically White names such as Emily Walsh and Greg Baker, and gave the other half of the applicants stereotypically Black names such as Lakisha Washington and Jamal Jones.

    6.45 percent of the applicants with Black-sounding names received a call back. What percent of the applicants with White-sounding names received a call back?

    (Please guess between 0 and 100.)

    \textit{Correct answer: 9.65 percent}
\end{itemize}

\vspace{-1.5mm}
\noindent \textit{Source linked on results page: \url{http://datacolada.org/wp-content/uploads/2015/04/bertrand_mullanaithan-1.pdf}}

\subsubsection*{Gun Deaths (Wave 1)}
\label{guns-question}
Participants randomly see one of the following questions:

\begin{itemize}
    \item After a mass shooting in 1996, Australia passed a massive gun control law called the National Firearms Agreement (NFA). The law illegalized, bought back, and destroyed almost one million firearms by 1997, mandated that all non-destroyed firearms be registered, and required a lengthy waiting period for firearm sales.

    Democrats and Republicans have each pointed to the NFA as evidence for/against stricter gun laws. This question asks about the effect of the NFA on the homicide rate in Australia.

    In the five years before the NFA (1991-1996), there were 319.8 homicides per year in Australia. In the five years after the NFA (1998-2003), how many homicides were there per year in Australia?

    \textit{Correct answer: 318.6}

    \item After a mass shooting in 1996, Australia passed a massive gun control law called the National Firearms Agreement (NFA). The law illegalized, bought back, and destroyed almost one million firearms by 1997, mandated that all non-destroyed firearms be registered, and required a lengthy waiting period for firearm sales.

    Democrats and Republicans have each pointed to the NFA as evidence for/against stricter gun laws. This question asks about the effect of the NFA on the homicide rate in Australia.

    In the five years after the NFA (1998-2003), there were 318.6 homicides per year in Australia. In the five years before the NFA (1991-1996), how many homicides were there per year in Australia?

    \textit{Correct answer: 319.8}
\end{itemize}

\vspace{-1.5mm}
\noindent \textit{Source linked on results page: \url{http://bit.ly/australia-homicide-rate}}

\subsubsection*{Violent Crime and Immigrants (Wave 1)}
\label{crime-immigrants-question}
Participants randomly see one of the following questions:

\begin{itemize}
    \item In 2015, German leader Angela Merkel announced an open-doors policy that allowed all Syrian refugees who had entered Europe to take up residence in Germany. From 2015-17, nearly one million Syrians moved to Germany. This question asks about the effect of Germany's open-doors refugee policy on violent crime rates.

    In 2017 (after the entrance of refugees), the violent crime rate in Germany was 228.2 per hundred-thousand people.
    
    In 2014 (before the influx of refugees), what was the violent crime rate in Germany per hundred-thousand people?

    \textit{Correct answer: 224.0 per hundred-thousand people}

    \item In 2015, German leader Angela Merkel announced an open-doors policy that allowed all Syrian refugees who had entered Europe to take up residence in Germany. From 2015-17, nearly one million Syrians moved to Germany. This question asks about the effect of Germany's open-doors refugee policy on violent crime rates.

    In 2014 (before the influx of refugees), the violent crime rate in Germany was 224.0 per hundred-thousand people.

    In 2017 (after the entrance of refugees), what was the violent crime rate in Germany per hundred-thousand people?

    \textit{Correct answer: 228.2 per hundred-thousand people}
\end{itemize}

\vspace{-1.5mm}
\noindent \textit{Source linked on results page: Main site: \url{bit.ly/germany-crime-main-site}. 2014-15 data: \url{bit.ly/germany-crime-2014-2015}. 2016-17 data: \url{bit.ly/germany-crime-2016-2017}.}

\subsubsection*{Violent Crime in the United States (Wave 1)}
\label{crime-us-question}
Participants randomly see one of the following questions:

\begin{itemize}
    \item This question asks how murder and manslaughter rates changed during the Obama administration. In 2008 (before Obama became president), the murder and manslaughter rate was 54 per million Americans.

    In 2016 (at the end of Obama's presidency), what was the per-million murder and manslaughter rate?

    \textit{Correct answer: 54 per million Americans}

    \item This question asks how murder and manslaughter rates changed during the Obama administration. In 2016 (at the end of Obama's presidency), the murder and manslaughter rate was 53 per million Americans.

    In 2008 (before Obama became president), what was the per-million murder and manslaughter rate?

    \textit{Correct answer: 54 per million Americans}
\end{itemize}

\vspace{-1.5mm}
\noindent \textit{Source linked on results page: \url{http://bit.ly/us-crime-rate}}

\vspace{-1mm}

\subsection*{Other Questions}

\subsubsection*{Performance}
\label{performance-question}
How well do you think you performed on this study about political and U.S. knowledge? I've compared the average points you scored for all questions (prior to this one) to that of 100 other participants.

How many of the 100 do you think you scored higher than?

(Please guess between 0 and 100.)

\subsubsection*{Latitude of Center of the United States (Wave 2)}
\label{latitude-question}
The U.S. National Geodetic Survey approximated the geographic center of the continental United States. (This excludes Alaska and Hawaii, and U.S. territories.)

How many degrees North is this geographic center?

(Please guess between 0 and 90. The continental U.S. lies in the Northern Hemisphere, the Equator is 0 degrees North, and the North Pole is 90 degrees North.)

\vspace{2mm}

\textit{Correct answer: 39.833.}

\textit{Source linked on results page: \url{http://bit.ly/center-of-the-us}}

\subsubsection*{Comprehension Check: Current Year (Waves 1+2)}
\label{comprehension-question}
In 1776 our fathers brought forth, upon this continent, a new nation, conceived in Liberty, and dedicated to the proposition that all men are created equal.

What is the year right now?

This is not a trick question and the first sentence is irrelevant; this is a comprehension check to make sure you are paying attention. For this question, your lower and upper bounds should be equal to your guess if you know what year it currently is.

\vspace{2mm}

\textit{Correct answer: 2019.}

\textit{Source linked on results page: \url{http://bit.ly/what-year-is-it}}

\section{Study Materials: Screenshots}

\subsection{Main Experiment}
\label{screenshots}

\begin{center}
\includegraphics[height = .9\textheight]{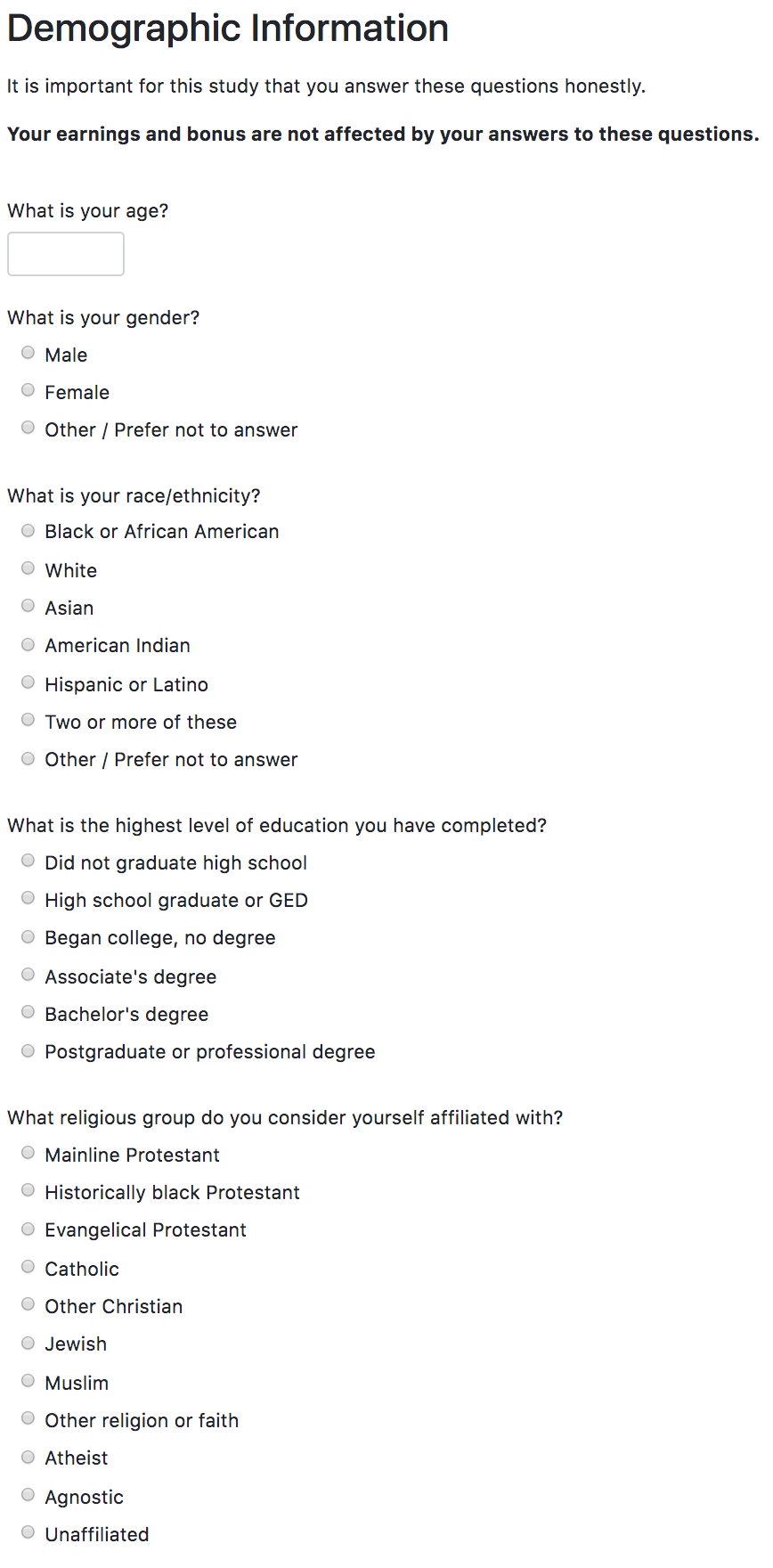}
\end{center}

\clearpage

\begin{center}
\includegraphics[width = \textwidth]{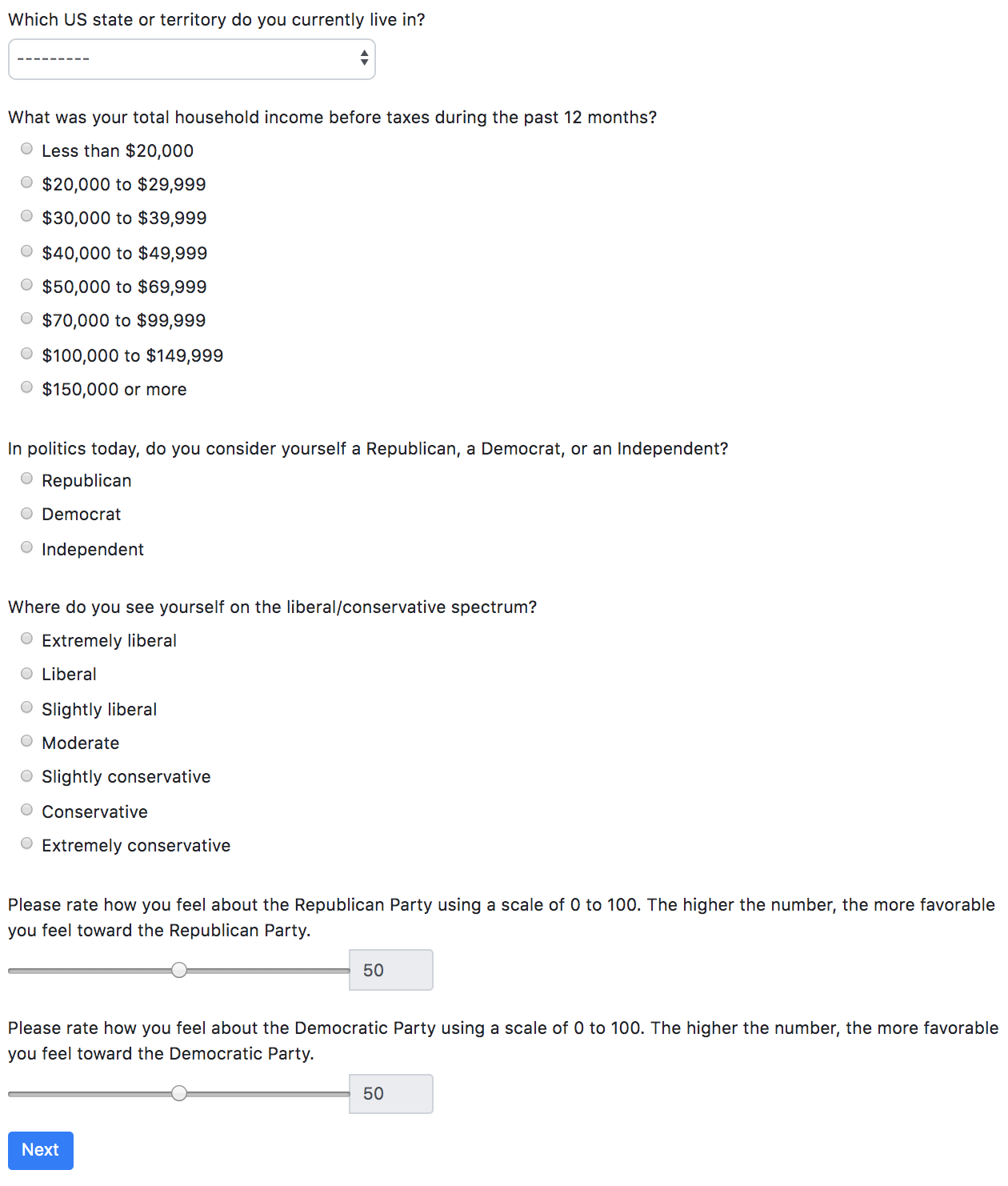}
\end{center}

\newpage

\begin{center}
\includegraphics[width = .95\textwidth]{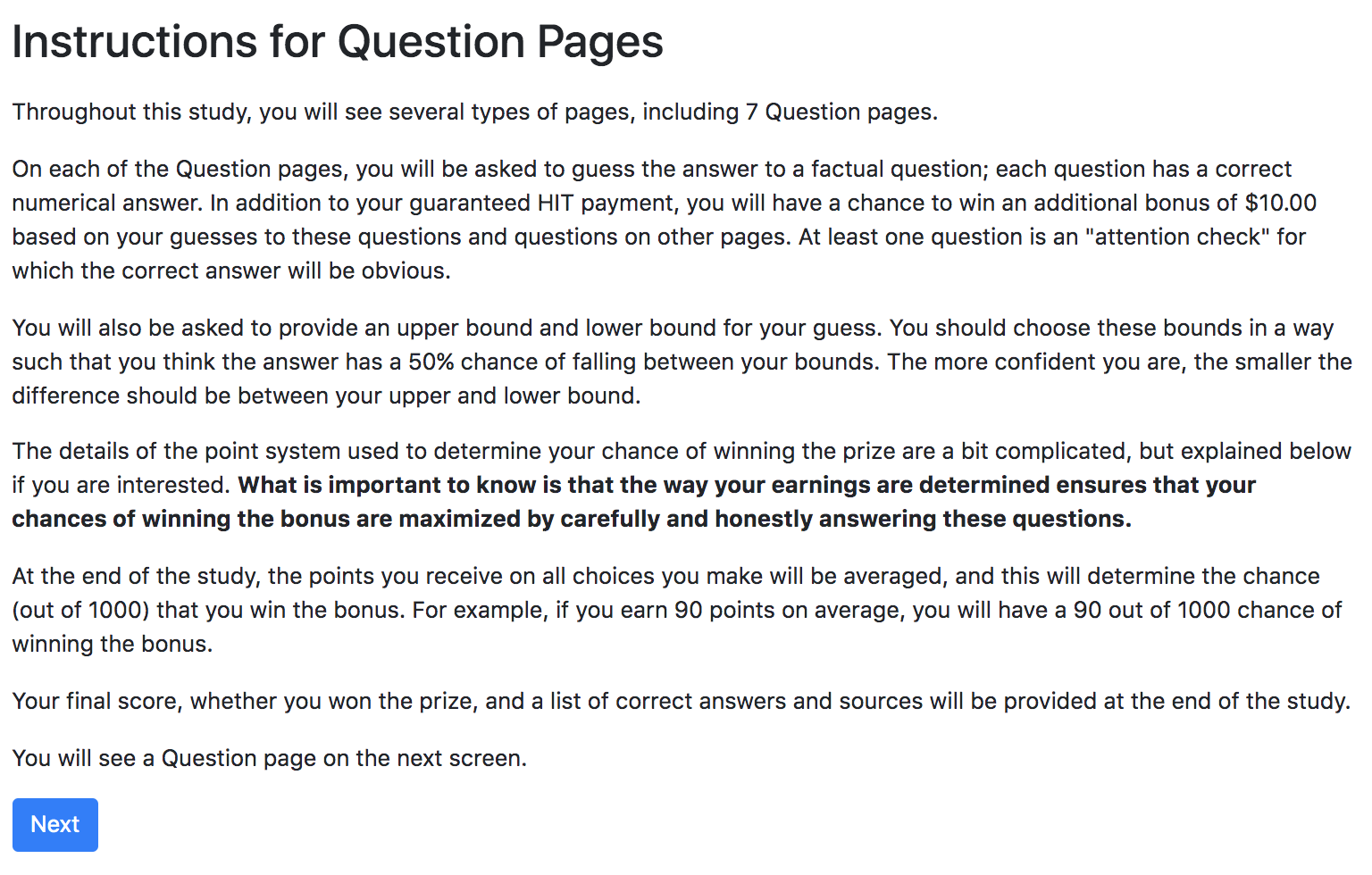}
\label{question-instructions}
\end{center}

\begin{center}
\includegraphics[width = \textwidth]{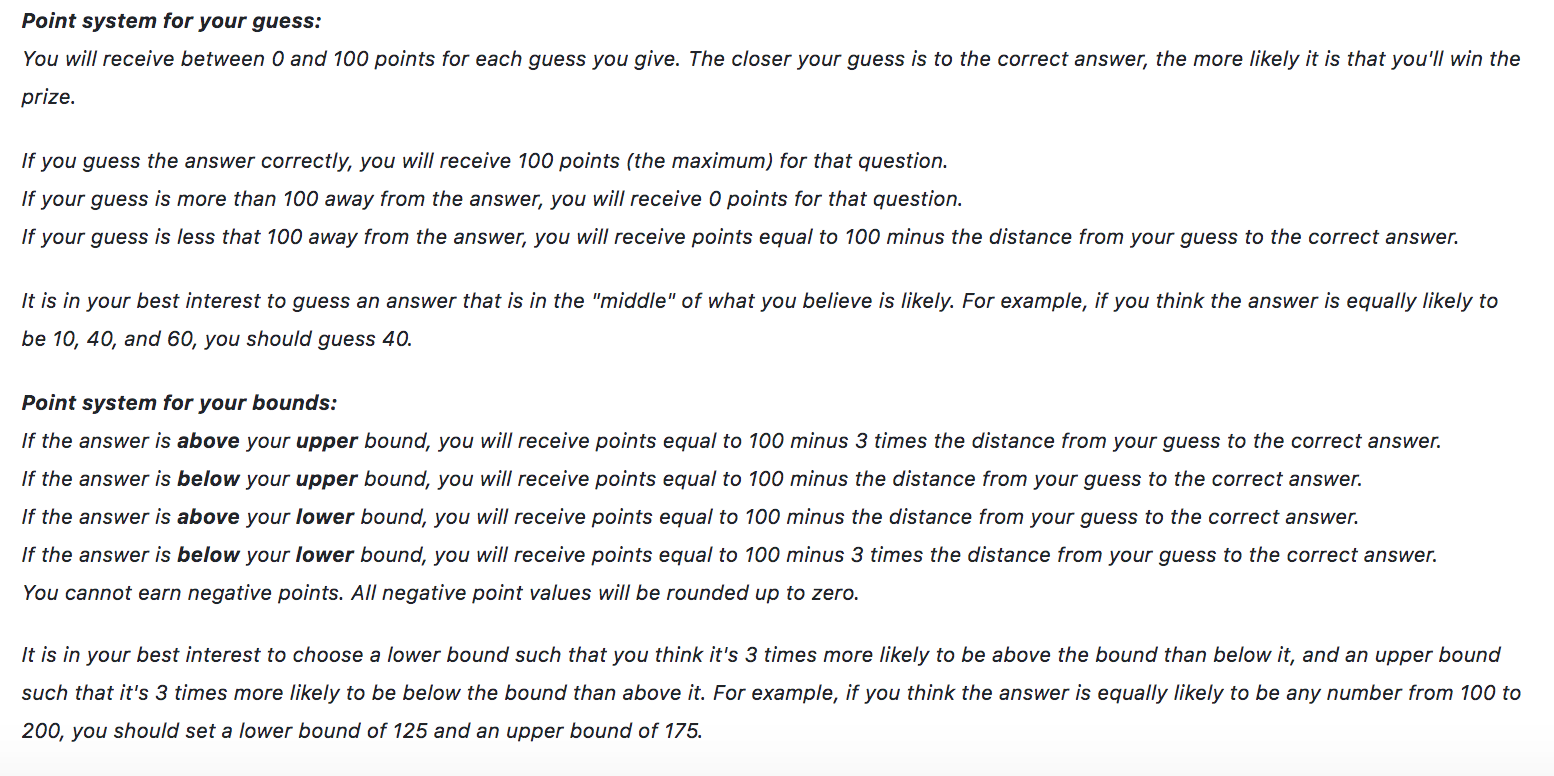}
\end{center}

\begin{center}
\includegraphics[width = \textwidth]{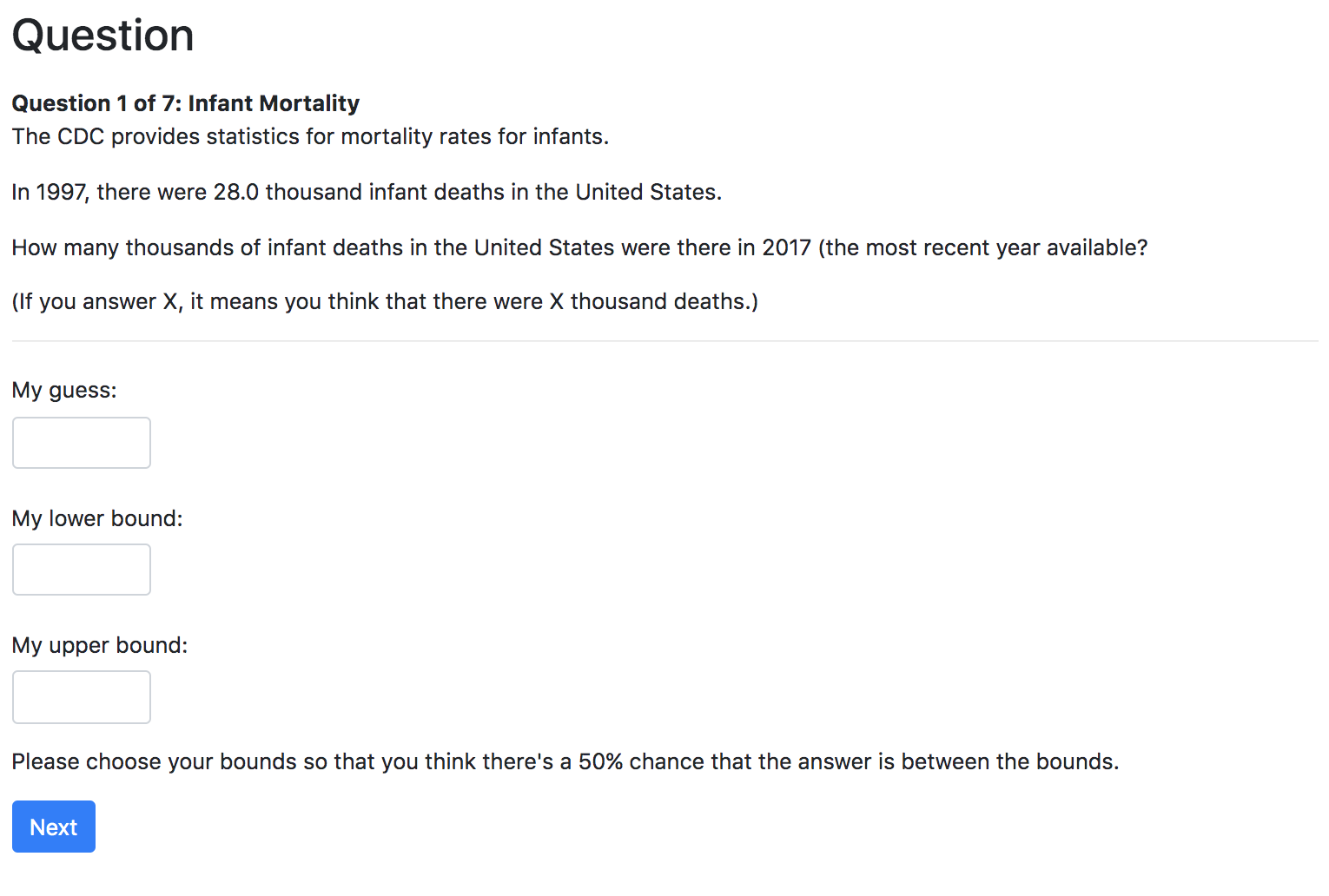}
\end{center}

\newpage

\begin{center}
\includegraphics[width = \textwidth]{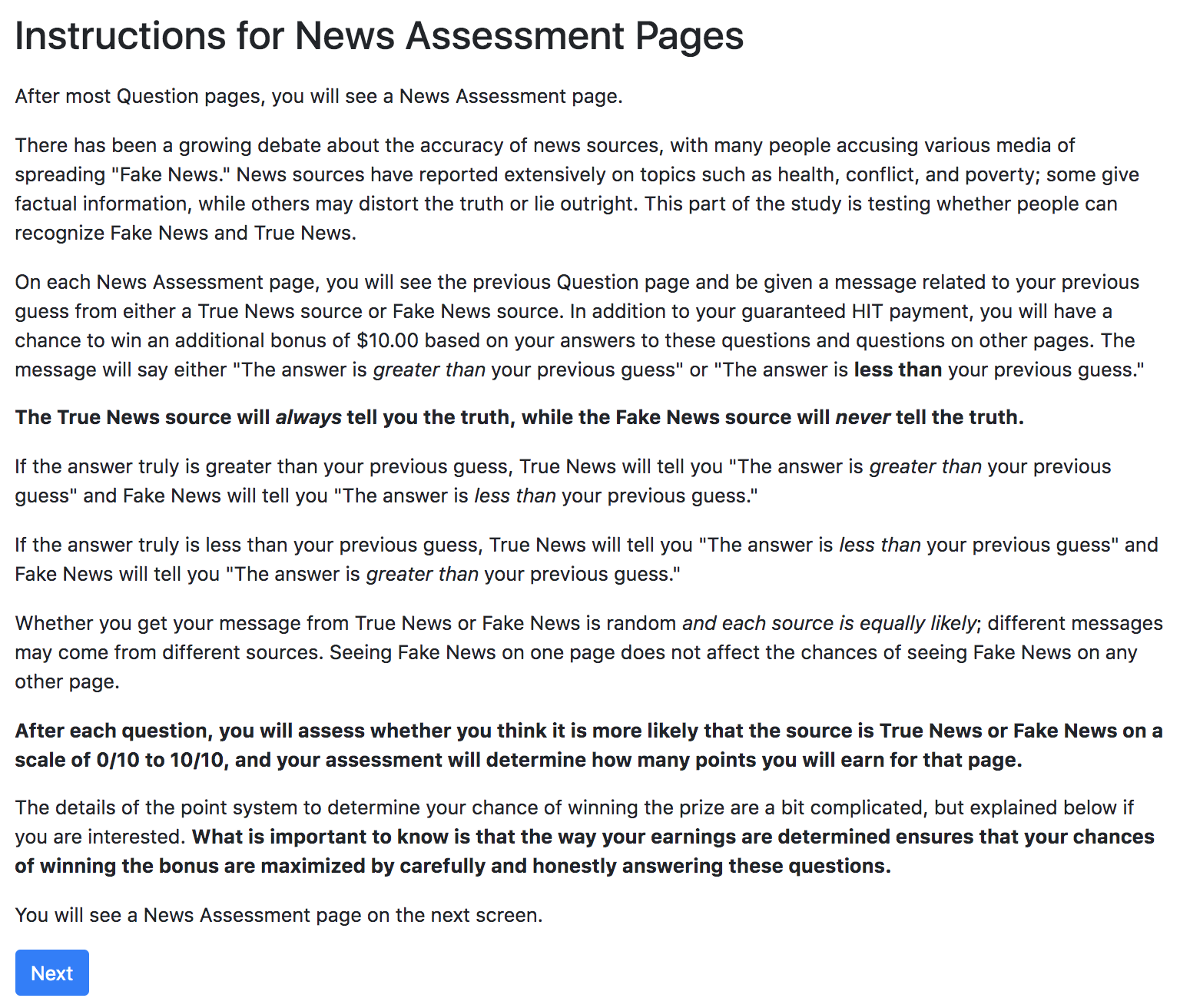}
\label{news-instructions}
\end{center}

\begin{center}
\includegraphics[width = \textwidth]{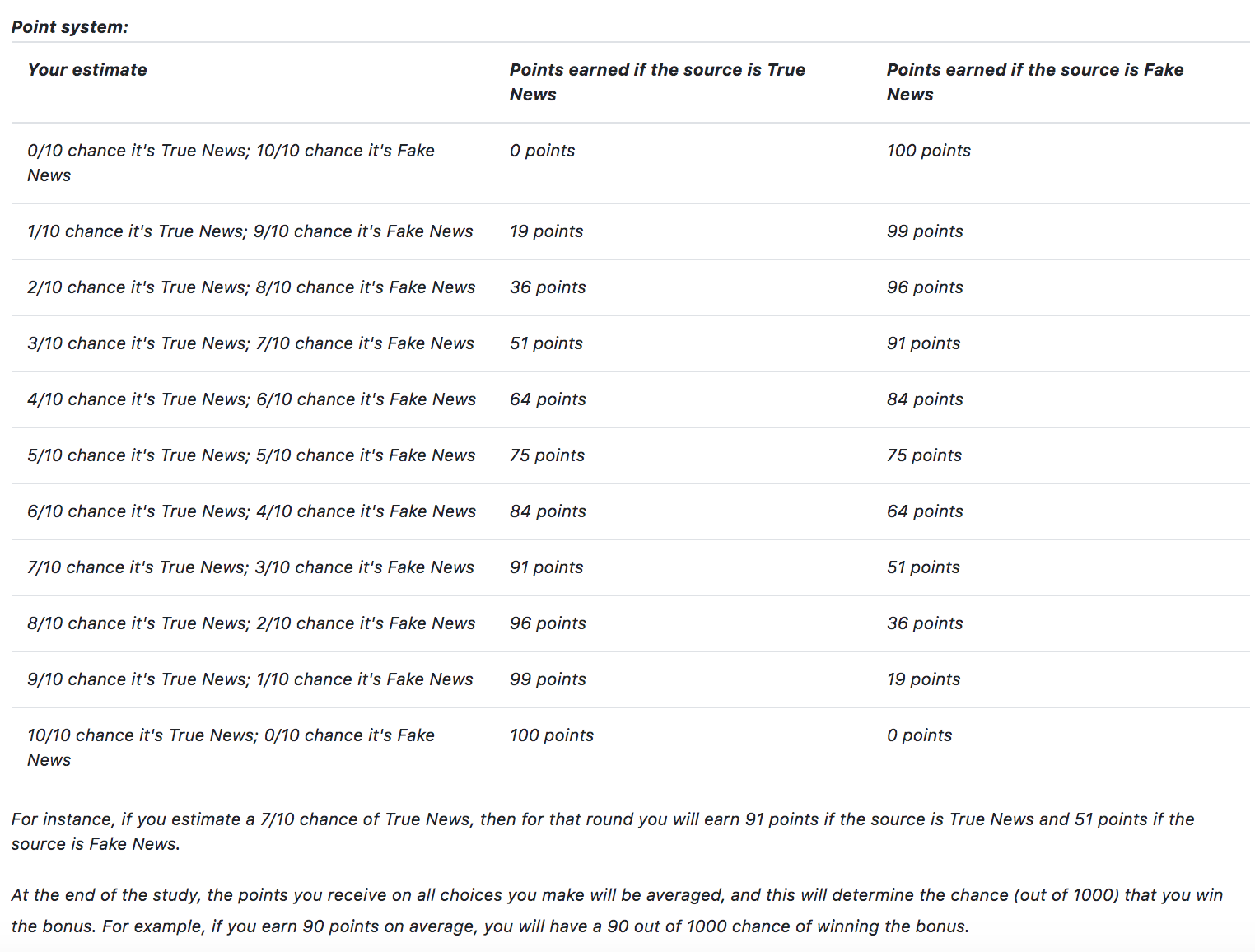}
\end{center}

\newpage

\begin{center}
\includegraphics[width = \textwidth]{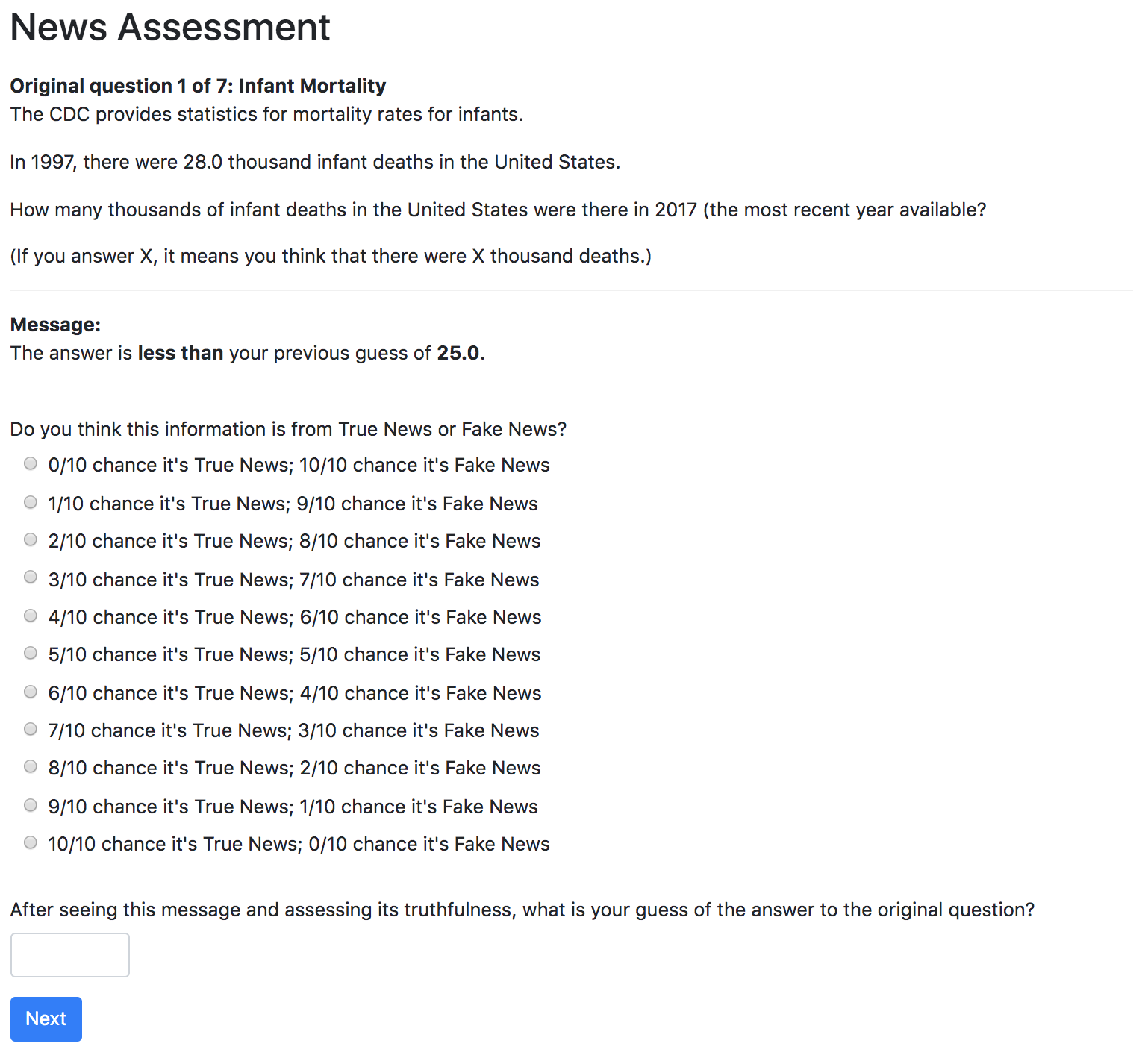}
\end{center}

\newpage 

\subsection{Survey Questions}

\begin{figure}[!htb]
\begin{center}
\includegraphics[width = .63\textwidth]{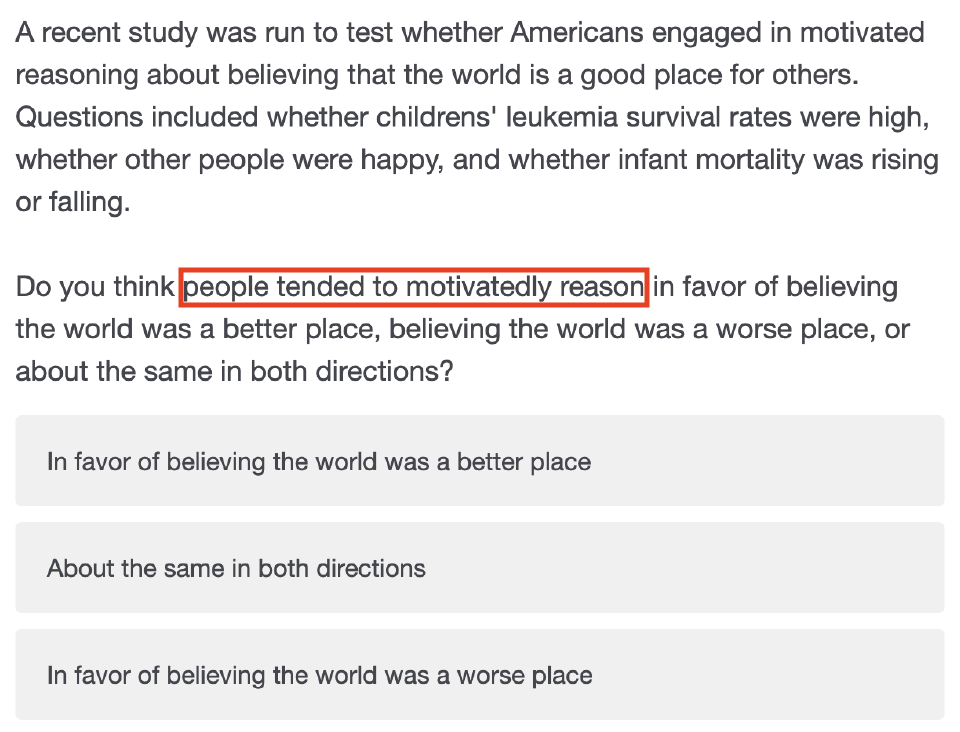}
\end{center}
\vspace{-6mm}
\caption{Survey 1: Good news and motivated reasoning. Red box not shown to participants.}
\label{screenshot-survey1}
\end{figure}

\vspace{-3mm}

\begin{figure}[!htb]
\begin{center}
\includegraphics[width = .63\textwidth]{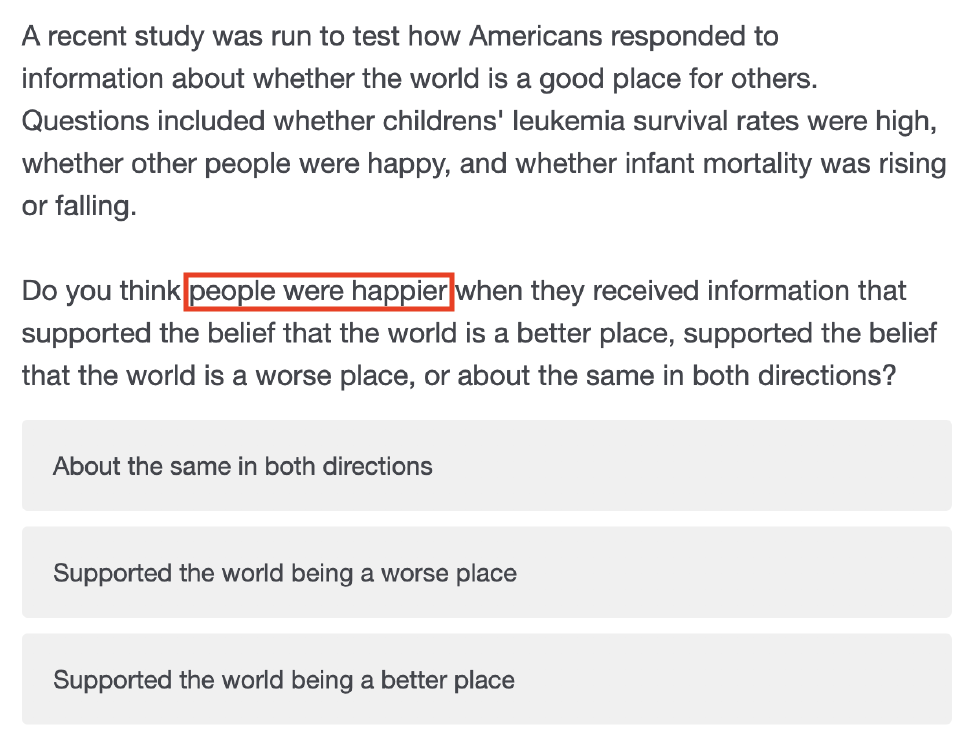}
\end{center}
\vspace{-6mm}
\caption{Survey 2: Good news and happiness. Red box not shown to participants.}
\label{screenshot-survey2}
\end{figure}





\end{document}